\documentclass[11pt,a4paper]{article} 
\pdfoutput=1
\usepackage{style}
\usepackage{epstopdf}
\usepackage{epsfig}
\usepackage{amsmath,amssymb,bm,amsfonts,yfonts
}
\newcommand{\be}{\begin{equation}}
\newcommand{\ee}{\end{equation}}
\newcommand{\bea}{\begin{eqnarray}}
\newcommand{\nn}{\nonumber}
\newcommand{\eea}{\end{eqnarray}}
\newcommand{\pd}{\partial}
\newcommand{\tf}{\tilde{f}}
\newcommand{\tth}{\tilde{\theta}}

\newcommand{\td}{\tilde{d}}

\newcommand{\bp}{\mathbf{p}}
\newcommand{\bq}{\mathbf{q}}
\newcommand{\bk}{\mathbf{k}}
\newcommand{\br}{\mathbf{r}}
\newcommand{\bx}{\mathbf{x}}

\def \lta {\mathrel{\vcenter
     {\hbox{$<$}\nointerlineskip\hbox{$\sim$}}}}
\def \gta {\mathrel{\vcenter
     {\hbox{$>$}\nointerlineskip\hbox{$\sim$}}}}

\newcommand{\lsim}{
\mathrel{\hbox{\rlap{\hbox{\lower4pt\hbox{$\sim$}}}\hbox{$<$}}}}

\newcommand{\gsim}{
\mathrel{\hbox{\rlap{\hbox{\lower4pt\hbox{$\sim$}}}\hbox{$>$}}}}

\DeclareSymbolFont{mathscrUC}{U}{rsfs}{m}{n}  
\DeclareSymbolFont{mathscrLC}{OT1}{pzc}{m}{n} 

\makeatletter
\DeclareRobustCommand*{\mathscr}[1]{\gdef\F@ntPrefix{mathscr@char@}%
  \@EachCharacter #1\@EndEachCharacter}
\long\def\DoLongFutureLet #1#2#3#4{%
   \def\@FutureLetDecide{#1#2\@FutureLetToken
      \def\@FutureLetNext{#3}\else
      \def\@FutureLetNext{#4}\fi\@FutureLetNext}
   \futurelet\@FutureLetToken\@FutureLetDecide}
\def\DoFutureLet #1#2#3#4{\DoLongFutureLet{#1}{#2}{#3}{#4}}
\def\@EachCharacter{\DoFutureLet{\ifx}{\@EndEachCharacter}%
   {\@EachCharacterDone}{\@PickUpTheCharacter}}
\def\m@keCharacter#1{\csname\F@ntPrefix#1\endcsname}
\def\@PickUpTheCharacter#1{\m@keCharacter{#1}\@EachCharacter}
\def\@EachCharacterDone \@EndEachCharacter{}

\DeclareMathSymbol{\mathscr@char@A}{\mathord}{mathscrUC}{`A}
\DeclareMathSymbol{\mathscr@char@B}{\mathord}{mathscrUC}{`B}
\DeclareMathSymbol{\mathscr@char@C}{\mathord}{mathscrUC}{`C}
\DeclareMathSymbol{\mathscr@char@D}{\mathord}{mathscrUC}{`D}
\DeclareMathSymbol{\mathscr@char@E}{\mathord}{mathscrUC}{`E}
\DeclareMathSymbol{\mathscr@char@F}{\mathord}{mathscrUC}{`F}
\DeclareMathSymbol{\mathscr@char@G}{\mathord}{mathscrUC}{`G}
\DeclareMathSymbol{\mathscr@char@H}{\mathord}{mathscrUC}{`H}
\DeclareMathSymbol{\mathscr@char@I}{\mathord}{mathscrUC}{`I}
\DeclareMathSymbol{\mathscr@char@J}{\mathord}{mathscrUC}{`J}
\DeclareMathSymbol{\mathscr@char@K}{\mathord}{mathscrUC}{`K}
\DeclareMathSymbol{\mathscr@char@L}{\mathord}{mathscrUC}{`L}
\DeclareMathSymbol{\mathscr@char@M}{\mathord}{mathscrUC}{`M}
\DeclareMathSymbol{\mathscr@char@N}{\mathord}{mathscrUC}{`N}
\DeclareMathSymbol{\mathscr@char@O}{\mathord}{mathscrUC}{`O}
\DeclareMathSymbol{\mathscr@char@P}{\mathord}{mathscrUC}{`P}
\DeclareMathSymbol{\mathscr@char@Q}{\mathord}{mathscrUC}{`Q}
\DeclareMathSymbol{\mathscr@char@R}{\mathord}{mathscrUC}{`R}
\DeclareMathSymbol{\mathscr@char@S}{\mathord}{mathscrUC}{`S}
\DeclareMathSymbol{\mathscr@char@T}{\mathord}{mathscrUC}{`T}
\DeclareMathSymbol{\mathscr@char@U}{\mathord}{mathscrUC}{`U}
\DeclareMathSymbol{\mathscr@char@V}{\mathord}{mathscrUC}{`V}
\DeclareMathSymbol{\mathscr@char@W}{\mathord}{mathscrUC}{`W}
\DeclareMathSymbol{\mathscr@char@X}{\mathord}{mathscrUC}{`X}
\DeclareMathSymbol{\mathscr@char@Y}{\mathord}{mathscrUC}{`Y}
\DeclareMathSymbol{\mathscr@char@Z}{\mathord}{mathscrUC}{`Z}
\DeclareMathSymbol{\mathscr@char@a}{\mathord}{mathscrLC}{`a}
\DeclareMathSymbol{\mathscr@char@b}{\mathord}{mathscrLC}{`b}
\DeclareMathSymbol{\mathscr@char@c}{\mathord}{mathscrLC}{`c}
\DeclareMathSymbol{\mathscr@char@d}{\mathord}{mathscrLC}{`d}
\DeclareMathSymbol{\mathscr@char@e}{\mathord}{mathscrLC}{`e}
\DeclareMathSymbol{\mathscr@char@f}{\mathord}{mathscrLC}{`f}
\DeclareMathSymbol{\mathscr@char@g}{\mathord}{mathscrLC}{`g}
\DeclareMathSymbol{\mathscr@char@h}{\mathord}{mathscrLC}{`h}
\DeclareMathSymbol{\mathscr@char@i}{\mathord}{mathscrLC}{`i}
\DeclareMathSymbol{\mathscr@char@j}{\mathord}{mathscrLC}{`j}
\DeclareMathSymbol{\mathscr@char@k}{\mathord}{mathscrLC}{`k}
\DeclareMathSymbol{\mathscr@char@l}{\mathord}{mathscrLC}{`l}
\DeclareMathSymbol{\mathscr@char@m}{\mathord}{mathscrLC}{`m}
\DeclareMathSymbol{\mathscr@char@n}{\mathord}{mathscrLC}{`n}
\DeclareMathSymbol{\mathscr@char@o}{\mathord}{mathscrLC}{`o}
\DeclareMathSymbol{\mathscr@char@p}{\mathord}{mathscrLC}{`p}
\DeclareMathSymbol{\mathscr@char@q}{\mathord}{mathscrLC}{`q}
\DeclareMathSymbol{\mathscr@char@r}{\mathord}{mathscrLC}{`r}
\DeclareMathSymbol{\mathscr@char@s}{\mathord}{mathscrLC}{`s}
\DeclareMathSymbol{\mathscr@char@t}{\mathord}{mathscrLC}{`t}
\DeclareMathSymbol{\mathscr@char@u}{\mathord}{mathscrLC}{`u}
\DeclareMathSymbol{\mathscr@char@v}{\mathord}{mathscrLC}{`v}
\DeclareMathSymbol{\mathscr@char@w}{\mathord}{mathscrLC}{`w}
\DeclareMathSymbol{\mathscr@char@x}{\mathord}{mathscrLC}{`x}
\DeclareMathSymbol{\mathscr@char@y}{\mathord}{mathscrLC}{`y}
\DeclareMathSymbol{\mathscr@char@z}{\mathord}{mathscrLC}{`z}
\makeatother

\relax

\title{Nonlinear evolution of density and flow perturbations on a Bjorken background}

\author[a]{Nikolaos Brouzakis} 
\author[b]{, Stefan Floerchinger}
\author[a,b]{, Nikolaos Tetradis}
\author[b]{, Urs Achim Wiedemann}

\affiliation[a]{Department of Physics, University of Athens, Zographou 157 84, Greece}
\affiliation[b]{Physics Department, Theory Unit, CERN, CH-1211 Gen\`eve 23, Switzerland}

\emailAdd{nbruzak@phys.uoa.gr}
\emailAdd{stefan.floerchinger@cern.ch}
\emailAdd{ntetrad@phys.uoa.gr}
\emailAdd{urs.wiedemann@cern.ch}

\abstract{
Density perturbations and their dynamic evolution from early to late times can be used for an improved understanding of interesting physical 
phenomena both in cosmology and in the context of heavy-ion collisions. We discuss the spectrum and bispectrum of these perturbations around a longitudinally expanding fireball after a heavy-ion collision. The time-evolution equations couple the spectrum and bispectrum to each other, 
as well as to higher-order correlation functions through nonlinear terms. A non-trivial bispectrum is thus always generated, even if absent initially. 
For initial conditions corresponding to a model of independent sources, we discuss the linear and nonlinear evolution is detail. We show that, if the initial conditions are sufficiently smooth for fluid dynamics to be applicable, the nonlinear effects are relatively small.}

\begin{document}

\maketitle

\section{Introduction}
It has been noted repeatedly~\cite{Mishra:2008dm,Mocsy:2010um,Naselsky:2012nw}
that the collective dynamics of the hot and dense QCD matter produced in ultra-relativistic heavy-ion 
collisions shares many commonalities with the expansion dynamics of the Early Universe. 
Both systems are initially relativistic, of high energy density and rapidly expanding.
Relativistic fluid dynamics describes at least some stages of their evolution, while particle species freeze-out from this fluid at characteristic densities. 
With the progress in measuring and analyzing fluctuations in the cosmic microwave background (CMB)~\cite{Hinshaw:2012aka,Ade:2013zuv}, 
and with the discovery that flow measurements in heavy-ion collisions result from the propagation of ``initial" event-by-event fluctuations \cite{Alver:2008zza,Alver:2010gr,Sorensen:2010zq},
it is by now clear that the little bangs studied in heavy-ion collisions and the Big Bang studied in cosmology correspond to the smallest and 
largest physical systems, respectively, for which fluctuation analyses can provide information about material properties. 
Here, we recall these commonalities in order to raise the prospect that progress in either field could be made by transferring 
techniques and insights from the other. There are, of course,  basic features with respect to which the two systems in question are 
quite distinct: for instance, both systems are characterized by vastly different scales and their dynamics is 
governed by different fundamental forces. On the other hand, there is an underlying unifying principle: the applicability of a relativistic fluid description.

In heavy-ion collisions, the nonlinearities in the evolution are expected to be maximal initially, 
but they decrease with time due to dissipative effects. 
In cosmology, the initial density perturbations are very small and linear perturbation theory is a good description at early times. 
But at late times, the gravitational instability leads to strong growth, resulting in the emergence of large-scale structure (LSS)
and the formation of galaxies and galaxy clusters. It is apparent then that the formulations of cosmological perturbation theory
most relevant for our purposes are the ones applied to the problem of LSS. 

Even though the focus, both in heavy-ion physics and in LSS cosmology, is on propagating perturbations  in fluid dynamics, 
there are marked differences in the way the dynamics of these perturbations is implemented. In heavy-ion physics, one usually initializes 
the fluid evolution with the help of Glauber Monte Carlo models or their variants based on saturation physics~\cite{Drescher:2007ax,Miller:2007ri,Alver:2008aq,Broniowski:2007nz}, which
specify the fluid dynamic fields including their fluctuations at the initial time.
Typically, these fields are then propagated numerically through hydro-solvers \cite{Bozek:2011ua,Schenke:2010rr,Vredevoogd:2012ui,DelZanna:2013eua}
up to the final state, in an effort to map out the relation between the (in principle wide) model space of conceivable initial conditions and the 
experimentally accessible data that can constrain them. Only very recently the first efforts were made to abstract from this explicit propagation 
of many model-dependent initial conditions, for example by characterizing the statistics of initial perturbations in terms of  
$n$-point correlation functions \cite{Bzdak:2013rya,Yan:2013laa,Floerchinger:2014fta}.
In cosmology on the other hand, the fluid dynamic equations of motion usually are not solved for individual fluctuating fields. Instead, the objects 
of interest are correlation functions, such as the two-point correlation function of fields (the spectrum), 
the three-point (the bispectrum) as well as higher-order correlation functions. 
The various analytical methods that have been developed 
in order to go beyond linear perturbation theory rely on  
resummations of subsets of perturbative diagrams  \cite{bernardeau0,CrSc1,CrSc2,taruya,taruya2,Max1,time,mcdonald,Blas1,blas,Manzotti,Wang,bernardeau}
or the use of effective field theory \cite{eff1,eff2,eff3}. 

A transparent and efficient approach to the problem of LSS in cosmology is the time-renormalization-group (time-RG) formulation \cite{time}. 
Within this approach, the physical system is described in terms of a hierarchy of 
coupled evolution equations for the spectrum, bispectrum and higher spectra. 
The method 
has been applied to ${\rm \Lambda CDM}$ (cosmological constant plus cold dark matter) and quintessence cosmologies \cite{time}, allowing for 
a possible coupling of dark energy to dark matter \cite{coupled}, or a variable equation of state \cite{wz}. It has also been
used for the study of models with neutrinos \cite{lesgourgues}.  
The main aim of the present paper is to apply this formulation of relativistic fluid dynamics commonly employed in cosmology to a simplified test 
system often studied in heavy-ion physics. More specifically, we shall apply the time-RG  
to the problem of time evolution of two- and three-point correlation functions of density and flow fluctuations on top of a Bjorken background \cite{bjorken}.

It must be noted that the issue of initial conditions is more complex in heavy-ion collisions: 
Immediately after the collision the local deviations from a smooth background are expected to be large, at least according to many 
current phenomenological models, in which the initial energy distribution is dominated by the ``lumpy'' substructure of 
nuclei in terms of their constituents (nucleons) and the spatial distribution of nucleon scattering centers. 
On the other hand, it is clear that fluid dynamics strictly applies only after a phase of early non-equilibrium evolution during which dissipative effects are 
very large and smoothen the spatial structures, at least to some extent. 
An interesting question is whether the fluid dynamics of heavy-ion collisions is essentially linear or strongly nonlinear 
at the time and length scales of a typical event and for those regimes of momenta (or wavevectors) for which 
the gradient expansion that underlies fluid dynamics is applicable.

This question may also depend on the kind of perturbations considered. For vorticity fluctuations, two of us have argued some time ago that the time and length scales of a heavy-ion collision might be sufficient in order to see the onset of nonlinear or turbulent behavior \cite{flowied}. 
On the other hand, within the current phenomenological picture
density or compressional modes (sound waves) are much more important. 
In the present paper we concentrate entirely on the latter and assume that the fluid velocity is vorticity free.

The main subject of the present paper is to discuss explicitly the time evolution of $n$-point correlation functions of fluid dynamic variables during the hydrodynamic regime of a heavy-ion collision. In particular, we derive and (approximately) solve differential equations directly for the time evolution of the correlation functions. This approach is different from the one of Teaney and Yan \cite{Teaney:2012ke}, in which the relation between the initial conditions (characterized by appropriately defined eccentricities) and observables (in particular the harmonic flow coefficients and correlations between event-plane angles) are formulated in terms of linear and non-linear response coefficients, but the latter are determined in a more indirect way via a numerical hydro-solver. The present approach also differs from ref.\ \cite{Floerchinger:2013tya} that used a hydro-solver to show that the fluid dynamic response to the initial conditions (characterized in this case my a Bessel-Fourier expansion) is close to linear, in the sense that a perturbative treatment becomes possible. 

This paper is organized as follows: we introduce the Bjorken model and the treatment of fluctuations in section~\ref{sec2}. In section~\ref{correl}, we derive the corresponding evolution
equations for two-point and three-point correlation functions of fluid dynamic fields and we discuss how they can be solved perturbatively. Section~\ref{sec:InitialConditions} introduces a model
for the initial conditions. The UV sensitivity of fluid dynamics determines its range of applicability and is discussed in detail in section~\ref{sec5} before we turn to the discussion of numerical results in section~\ref{sec6}. The main results of our study are then summarized in the conclusions.
All dimensionful parameters are measured in femtometers (fm) when units are not explicitly indicated. 

\section{Fluid dynamics of perturbations in the Bjorken model}
\label{sec2}

%
We start from the relativistic hydrodynamic equations of motion for a fluid without any conserved charges. 
Keeping the effects of shear viscosity ($\eta$), but not of bulk viscosity, within the first order in a gradient expansion, 
the equations of motion read
\bea
u^\mu \nabla_\mu \epsilon + (\epsilon + p) \nabla_{\mu} u^\mu - 2 \,\eta \,\sigma_{\mu\nu} \sigma^{\mu\nu} &=& 0\, ,
\label{eq:firstorderhyrdoepsilon} \\
(\epsilon+p) u^\mu \nabla_\mu  u^{\nu} + \Delta^{\nu\mu} \nabla_{\mu} p - 2 \,\eta\, \Delta^{\nu}_{\;\;\rho}\nabla_{\mu} \sigma^{\mu\rho} &=& 0\, .
\label{eq:firstorderhydrou}
\eea
Here, $\epsilon$ and $p$ denote the local energy density and pressure, respectively.
We work with metrics of signature $(-,+,+,+)$. In terms of the covariant derivative $\nabla_{\mu}$ and the projector
$\Delta^{\mu\nu} = g^{\mu\nu} + u^{\mu} u^{\nu}$ on the subspace orthogonal to the flow vector $u^\mu$, the
traceless and transverse (i.e. orthogonal to $u^\mu$) tensor $\sigma^{\mu\nu}$ takes the form
\begin{equation}
\sigma^{\mu\nu} = \frac{1}{2}(\Delta^{\mu}_{\;\;\rho} \nabla^{\rho} u^{\nu} + \Delta^{\nu}_{\;\;\rho} \nabla^{\rho} u^{\mu})-\frac{1}{3}\Delta^{\mu\nu}(\nabla_{\rho}u^{\rho}).
\end{equation}

The solution of eq. (\ref{eq:firstorderhydrou}) in full generality is a formidable (numerical) task. Instead, we shall employ in the following
a split between background and fluctuations in order to discuss a class of interesting solutions. More precisely, we start from an analytically solvable 
smooth {\it background} solution of eq. (\ref{eq:firstorderhydrou}) that was first formulated by Bjorken~\cite{bjorken} and captures,
despite its simplicity, phenomenologically relevant features. We then discuss the propagation of small perturbations
on top of this background solution. This approach is analogous to the one employed in cosmology, where fluctuations
propagate on top of the smooth Hubble flow of a Friedmann-Robertson-Walker (FRW) Universe. 

Bjorken's model~\cite{bjorken} is based on the idea that in nuclear collisions at ultra-relativistic energy particle production 
is almost flat in rapidity. This motivates the assumption that all fluid dynamic fields are initially independent of space-time
rapidity, $y=\text{arctanh}(x_3/|x_0|)$. Fluid dynamic evolution preserves this symmetry. Using the coordinates 
$(\tau, x_1, x_2, y)$, with proper time $\tau=\sqrt{x_0^{2}-x_3^{2}}$ and metric $g_{\mu\nu} =\text{diag}(-1,1,$ $1,\tau^2)$,
the solutions are then $y$-independent and the problem reduces to a (2+1)-dimensional one. A strong further simplification
is obtained if one also assumes translational and rotational invariance of the initial conditions in the transverse plane. This yields a
one-dimensional toy model, the so-called Bjorken model~\cite{bjorken}. In terms of the enthalpy $w=\epsilon+p = sT$ and the kinematic
viscosity $ \nu = \eta/w$,  the evolution equation 
of this one-dimensional model reads  
\begin{equation}
\partial_{\tau} \epsilon + \frac{w}{\tau}\left(1-\frac{4\nu}{3\tau}\right)= 0\, .
\label{eq:evolutioneps2}
\end{equation}
For $\nu/\tau \ll 1$ and an ideal equation of state $\epsilon = 3 p$, we obtain
the characteristic time dependencies of the Bjorken background for the energy density
\begin{equation}
\epsilon_\text{Bj}(\tau) = \epsilon_\text{Bj}(\tau_0) \left(\frac{\tau_0}{\tau}\right)^{4/3}\, ,
\label{eq:BjorkenEnergydensity}
\end{equation}
or, equivalently, the characteristic time-dependence of the Bjorken background temperature,
$ T_\text{Bj}(\tau) = T_\text{Bj}(\tau_0) \left({\tau_0}/{\tau}\right)^{1/3}$. For a time-independent 
normalized viscosity $\eta/s$, the ratio ${\nu_\text{Bj}(\tau)}/{\tau}$ in  eq.\ \eqref{eq:evolutioneps2}
decreases like 
\begin{equation}
\frac{\nu_\text{Bj}(\tau)}{\tau} 
= \left( \frac{\eta}{s} \right) \frac{1}{T_\text{Bj}(\tau_0)\, \tau_0} \left(\frac{\tau_0}{\tau}\right)^{2/3}\, .
\label{eq:Bjorkennutau}
\end{equation}
Therefore, replacing the bracket in eq.\ \eqref{eq:evolutioneps2} by unity is an
approximation that is consistent with the late time behavior.
This approximation is also justified at early times ($\tau_0\simeq 1$ fm, $T_\text{Bj}(\tau_0)\simeq 1/$fm) for small values
of the viscosity to entropy ratio, such as those favored by holography~\cite{Kovtun:2004de}: $\eta/s \simeq 1/(4\pi)$.

\subsection{Perturbations on a Bjorken background}
The propagation of perturbations on top of the Bjorken background solution has been discussed in ref.~\cite{flowied}.
Here, we recall the results relevant for the present work, and the approximations on which they are based, but we
refer to ref.~\cite{flowied} for a detailed derivation. The background solution is defined by (\ref{eq:BjorkenEnergydensity})
 and by the rapidity-independent Bjorken flow field $u_{\rm Bj}^{\mu} = \left(1,0,0,0 \right)$. We use an ideal equation of
state, $\epsilon = 3\, p$ and allow for small
local fluctuations around this solution. For the flow field, local fluctuations are parameterized by the transverse and rapidity
components $u^1, u^2, u^y$, respectively.  The normalization condition $u^\mu u_\mu = -1$
of the local fluid velocity $u^\mu=(u^\tau, u^1,u^2,u^y)$ implies then 
\begin{equation}
(u^\tau)^2 = 1 + (u^1)^2+ (u^2)^2 + \tau^2 (u^y)^2 = 1 + u_j u^j.
\label{eq:normalizationutau}
\end{equation}
Here and in the following, we use for latin indices a summation convention that runs over the spatial components $1,2,y$,
with the spatial part of the metric being $g_{ij} = \text{diag}(1,1,\tau^2)$. We consider small local perturbations in the sense that $u_ju^j(x) \ll 1$,
and we neglect terms that are parametrically suppressed due to $u_j u^j(x) \ll 1$ or $\nu/\tau \ll 1$ 
compared to other terms with the same combination of derivatives acting on the fields. For every combination of derivatives 
there is one term of lowest order which is {\it not} neglected. 
It is then straightforward to derive from eqs. (\ref{eq:firstorderhyrdoepsilon})
and (\ref{eq:firstorderhydrou}) equations of motion for the fluctuations $u^1, u^2, u^y$ in flow velocity, and for fluctuations
$\delta\epsilon$ in the energy density (see ref.~\cite{flowied}). These equations of motion take a particularly transparent form if they
are written in terms of a rescaled time variable
\begin{equation}
t = \frac{3 \,\tau^{4/3}}{4\, \tau_0^{1/3}}, \quad \partial_t \equiv \left(\frac{\tau_0}{\tau}\right)^{1/3} \partial_\tau\, ,
\label{eq:deft}
\end{equation}
and in terms of the logarithm of the perturbed energy density or, equivalently, temperature:
\begin{equation}
 d \equiv \ln\left(\frac{T}{T_\text{Bj}(\tau)}\right) =
\frac{1}{4}\ln\left(1+\frac{\delta \epsilon}{\epsilon_\text{Bj}}\right) \approx \frac{\delta \epsilon}{4\, \epsilon_\text{Bj}}\, .
\label{eq3.18}
\end{equation}
Here and in the following, we assume that  the perturbation $\delta \epsilon$ is small compared to the energy density $\epsilon_\text{Bj}$ of the background. Whether this assumption is justified depends of course on the initial conditions for the perturbation $\delta \epsilon$ and, to some extent, on the precise definition of the background energy density $\epsilon_\text{Bj}$. As discussed in section 6.2, the relative size of non-linear higher-order corrections can be used to assess a posteriori the validity of this perturbative approach for any fluctuation that one may want to consider. For another formulation of fluctuations in heavy-ion collisions~\cite{Floerchinger:2013tya}, it is known that a perturbative treatment remains valid up to remarkably large fluctuation amplitudes (fluctuations of the order of the background density). While there is no general statement that this needs to be the case, we anticipate that our results in section 6.2 are consistent with this earlier finding.
%

After the rescaling of the fluid dynamic fields,
\begin{equation}
u_j = \left(\frac{\tau}{\tau_0}\right)^{1/3} v_j\, ,\qquad d_{\tau}\equiv \left(\frac{\tau_0}{\tau}\right)^{2/3}\, d\, ,
\qquad \nu_0 = \nu \left(\frac{\tau_0}{\tau}\right)^{1/3}\, ,
\label{eq:rescaling}
\end{equation}
the equation of motion for the rescaled logarithm $d_{\tau}$ of the energy density takes the form
\begin{equation}
\begin{split}
 \partial_t d_{\tau}& + \frac{1}{2 t} d_{\tau} + \sum_{m=1}^2 v_m \partial_m d_{\tau} + \frac{1}{\tau^2} v_y \partial_y d_{\tau} + \frac{1}{3} \left( \frac{\tau_0}{\tau} \right)^{2/3} \partial_j\, v^j   \\
& - \frac{\nu_0}{6} {\Bigg [} \sum_{m,n=1}^2 (\partial_m v_n + \partial_n v_m)(\partial_m v_n+\partial_n v_m)
 + \frac{2}{\tau^2}\sum_{m=1}^2 (\partial_y v_m+\partial_m v_y)(\partial_y v_m+\partial_m v_y)\\
& ~~~~~~~~~~~~~~~~~~~~~~~~~~~~~~~~~~~~~~~~~~~~~~~~~~
+ \frac{4}{\tau^4} (\partial_y v_y)^2 -\frac{4}{3} \left(\partial_j\, v^j \right)^2
{\Bigg ]}=0\, ,
\end{split}
\label{eq:deqtransformed}
\end{equation}
and the equations of motion for the rescaled velocity components ($j=1,2,y$) read 
\begin{equation}
 \partial_t v_j + \sum_{m=1}^2 v_m \partial_m v_j + \frac{1}{\tau^2} v_y \partial_y v_j + \partial_j d_{\tau} -\frac{1}{3} \left(\partial_i\, v^i \right) \,v_j
 -\nu_0 \left[ \frac{1}{3} \partial_j \partial_i v^i + (\partial_1^2+\partial_2^2+\frac{1}{\tau^2} \partial_y^2) v_j \right] =0\, .
\label{eq:velocityeqtransformed}
\end{equation}
In the following, we denote the velocity divergence, sometimes also referred to as expansion scalar, by
\begin{equation}
\partial_j\, v^j= \partial_1 v_1 +\partial_2 v_2 + \frac{1}{\tau^2} \partial_y v_y \equiv \vartheta.
\label{eq:defvartheta}
\end{equation}

The two equations (\ref{eq:velocityeqtransformed}), (\ref{eq:defvartheta})
are the starting point of the present work. For the following, we subject them to the following additional approximations:
\begin{itemize}
	\item {\it We neglect fluctuations in the longitudinal $y$-direction.}\\
	   Remarkably, in the equations of motion (\ref{eq:deqtransformed}), (\ref{eq:velocityeqtransformed}), contributions from the $y$-dependence
	   of fluctuations are suppressed by powers of $1/\tau$. At late times, the equations of motion thus become effectively (2+1)-dimensional.
	   In the following, we make the assumption that the fluctuating initial conditions are independent of $y$. As a consequence, we deal 
 	   with a (2+1)-dimensional system at all times, and we consider equations of motion for $v_j=v_j(t,x_1,x_2)$, with $j=1,2$, but
	   not with $j=y$. 
       \item {\it We assume that the fluid is irrotational.}\\
	   In a perturbative treatment, fluctuations in vorticity couple only nonlinearly to density perturbations. Since a 
	   perturbative treatment of fluctuations is expected to be valid for the short time scale ($\lta 10$ fm) relevant for heavy-ion collisions,
           we neglect in the following vorticity and thus the effects of fluid turbulence. For an irrotational fluid, the velocity can then be written
	  as the gradient of a scalar: $v_m=\pd_m f$, with $m=1,2$. Taking the divergence of eq. (\ref{eq:velocityeqtransformed}),
	  one can write an evolution equation for the scalar $f$,
\bea \pd_m \pd_m \dot{f}&+&\left(\pd_m\pd_n f \right)\left(\pd_n \pd_m f \right)+\left(\pd_n f \right)
\left(\pd_m \pd_n \pd_m f \right)+\pd_m\pd_md_{\tau}
\nn \\
&-&\frac{1}{3} \left(\pd_m \pd_n \pd_n f \right) \pd_m f
-\frac{1}{3}\left( \pd_n \pd_n f \right)\left( \pd_m \pd_m f \right)-\frac{4 \nu_0}{3} \pd_m \pd_m \pd_n \pd_n f=0.
\label{eqff} \eea
A dot denotes a derivative   with respect to $t$, while repeated indices are summed. We have not used upper indices in order
to avoid confusion with the more general, rapidity-dependent case discussed earlier.
	\item {\it An IR cutoff can retain important aspect of a finite transverse extension.}\\
	The treatment of fluctuations on top of the Bjorken background is based on assuming their statistical homogeneity and 
 	isotropy in the transverse plane. This idealization can be meaningful only for distances smaller than the typical transverse extension 
	of a heavy-ion collision, which we take to be around 8 fm. It implies that the power spectrum depends only on the magnitude of the 
	momentum vector, while the bispectrum depends on the magnitudes of three momentum vectors.
	It is a severe approximation, since it prevents for instance the development
        of a transverse background flow, which would appear inevitably for systems of finite transverse extension. 
 	 Despite this limitation, 
	an important effect of the finiteness of the system can be retained in our treatment. Namely, the initial power spectrum is expected 
         to vanish below the momentum scale that corresponds to distances larger than $\sim8$ fm, as no 
	correlations within a beam are expected at such length scales. In principle, this can be implemented 
         by imposing an effective IR cut-off on the initial fluctuations. In practice, the results presented in the following are insensitive
        to this kind of IR cut-off. 
\end{itemize}

%
%

\section{Correlations and spectra}
\label{correl}
In the following we discuss fluctuations on top of a Bjorken background in a set-up akin to recent treatments of cosmological fluctuations. 
The role played by Friedmann's equation for the scale factor in the cosmological model will be played by 
the evolution equation for the Bjorken-expanding background fluid derived in section~\ref{sec2}. In order to discuss fluctuations around
the Bjorken solution similarly to cosmology, we first Fourier transform their evolution equations and derive the
corresponding equations for correlations and spectra.

\subsection{Nonlinear evolution equations in Fourier space}
As summarized in section~\ref{sec2}, equations  (\ref{eq:deqtransformed}), (\ref{eq:velocityeqtransformed}) are
our starting point for an analysis of the nonlinear evolution of transverse fluctuations induced by mode coupling and 
the leading  dissipative effects of viscosity. Here, we consider the case of an irrotational fluid where  
$v_m=\pd_m f$ for $m=1,2$, $v_y=0$, and $d_\tau$, $f$ functions of $t,x_1,x_2$. The primary dynamic fields
$d_{\tau}$ and $f$ can then be written in Fourier space as
\bea 
d_{\tau}(t,\bx)&=& \int d^2k\,\, e^{i\bk\bx} \td_{\bk}(t)\, ,
\nn \\
f(t,\bx)&=&\int d^2k \,\, e^{i\bk\bx}\tf_\bk(t)\, .
\label{fts} \eea
They satisfy the equations of motion
\bea
\dot{\td}_{\bk}&=&-\frac{1}{2t} \td_{\bk}+  \left(\frac{\tau_0}{12 t} \right)^{\frac{1}{2}}    k^2\tf_{\bk}
+\int d^2p\,  d^2q \, \delta(\bk-\bq-\bp)\, \alpha_1(\bp,\bq) \td_{\bp} \tf_{\bq} 
\nn \\
&&+\int d^2p \,  d^2q \, \delta(\bk-\bq-\bp) \, \alpha_2(\bp,\bq) \tf_{\bp} \tf_{\bq}\, ,
\label{fouf}\\
\dot{\tf}_{\bk}&=&-\frac{4 \nu_0}{3}k^2 \tf_\bk-\td_\bk+
\int d^2p\, d^2q \, \delta(\bk-\bq-\bp) \, \beta(\bp,\bq)  \tf_{\bp}  \tf_{\bq}\, ,
\label{foud}
\eea
where
\bea
\alpha_1(\bp,\bq)&=&\bp \bq \, ,
\label{alpha1} \\
\alpha_2(\bp,\bq)&=&\frac{2 \nu_0}{3} \left(\left(\bp \bq \right)^2-\frac{1}{3}p^2q^2\right)\, ,
\label{alpha2} \\ 
 \beta(\bp,\bq)&=&\frac{1}{(\bp+\bq)^2}\left[ \left(\bp\bq \right)^2-\frac{1}{3}p^2q^2+\frac{1}{3} \left(\bp \bq \right) \left(q^2+p^2\right)
\right]\, .
\label{beta} \eea 
We define the combined field $\phi_a(t,\bk)$, $a=1,2$ with
\be
\phi_a(t,\bk) \equiv \left(
\begin{array}{c}
\phi_1
\\
\phi_2
\end{array} \right)
=
\left(
\begin{array}{c}
\td_\bk
\\
\tf_\bk
\end{array} \right)\, ,
\ee
 and the matrix 
\be
\Omega(\bk,t)=\left(
\begin{array}{cc}
\frac{1}{2t} &~- \left(\frac{\tau_0}{12 t} \right)^{\frac{1}{2}} k^2
\\
 1 &  \frac{4 \nu_0}{3}k^2 
\end{array} \right).
\label{ome} \ee
Then, eqs. (\ref{fouf}), (\ref{foud}) take the form
\be
\dot\phi_a (\bk)= -\Omega_{ab}(\bk) \phi_b (\bk)+\int d^2p \, d^2q \, \delta(\bk-\bp-\bq)
\gamma_{abc}(\bp,\bq)\phi_b (\bp) \phi_c(\bq),
\label{eom}
\ee
where we have not indicated explicitly the dependence of $\phi$ on time. The nonzero elements of $\gamma_{abc}$ are 
\bea
\gamma_{121}(\bp,\bq)&=&\gamma_{112}(\bp,\bq)=\frac{\alpha_1(\bp,\bq)}{2}\, , \nn \\
\gamma_{122}(\bp,\bq)&=&\alpha_2(\bp,\bq)\, , \nn \\
\gamma_{222}(\bp,\bq)&=&\beta(\bp,\bq)\, .
\label{gammas} \eea 
Note that eq.\ \eqref{eom} implies that one can always define $\gamma_{abc}(\bp, \bq)$ such that
\begin{equation}
\gamma_{abc}(\bp,\bq) = \gamma_{acb}(\bq,\bp).
\label{eq:symmetryGamma}
\end{equation}
Moreover, because $d_\tau(t,\bx)\in\mathbb{R}$ and $f(t,\bx)\in\mathbb{R}$ are real fields one has $\phi_a(t,\bk) = \phi_a(t,-\bk)^*$ which implies 
\begin{equation}
\Omega_{ab}(\bk) = \Omega_{ab}(-\bk)^*, \quad\quad\quad
\gamma_{abc}(\bp,\bq) = \gamma_{abc}(-\bp,-\bq)^*.
\label{eq:realityconstraintOmegaGamma}
\end{equation}
The invariance of the time evolution equation \eqref{eom} under parity transformations $\phi_a(t,\bk) \to \phi_a(t,-\bk)$ (both components of $\phi_a$ transform as scalars) implies
\begin{equation}
\Omega_{ab}(\bk) = \Omega_{ab}(-\bk), \quad\quad\quad
\gamma_{abc}(\bp,\bq) = \gamma_{abc}(-\bp,-\bq).
\label{eq:parityconstraintOmegaGamma}
\end{equation}
 In addition, one observes from eqs.\ \eqref{gammas}, \eqref{alpha1}, \eqref{alpha2} and \eqref{beta} that 
\be
\gamma_{abc}(\bp,\bq)=\gamma_{abc}(\bq,\bp)=\gamma_{acb}(\bp,\bq).
\label{gamsym} \ee

\subsection{Definitions of expectation values and spectra}
\label{sec:defExpValuesSpectra}

For symmetry reasons only the zero mode with $\bk = 0$ can have an expectation value and we define
\be
\langle \phi_a(\bk) \rangle = \bar \phi_a \; \delta(\bk),
\label{def:expvalue}
\ee
Note that, because of the definition $v_m = \partial_m f$, a spatially constant part of $f(t,\bx)$ has no physical meaning. 
One can always change it without changing the velocity in order to achieve $\tilde f_{\bk = 0} = 0$ and therefore $\bar\phi_2=0$. 
On the other hand, a spatially constant part in $d_\tau(t,\bx)$ corresponds through eq.\ \eqref{eq3.18} 
to a deviation from the Bjorken background solution of eqs. 
(\ref{eq:BjorkenEnergydensity}), (\ref{eq:Bjorkennutau}). 
This is a physical effect that may arise when energy from the dissipation of perturbations heats up the system. 

The power spectrum $P_{ab}(\bk)$ is defined to correspond to the connected part of the two-point function
\be 
\langle \phi_a(\bk) \phi_b (\bp) \rangle=P_{ab}(\bk) \, \delta(\bk+\bp) + \bar\phi_a  \, \bar\phi_b \, \delta(\bk)  \delta(\bp) \, ,
\label{sp} \ee 
the bispectrum $B_{abc}(\bk,\bp,\bq)$ is defined by
\be 
\begin{split}
\langle \phi_a(\bk) \phi_b (\bp) \phi_c(\bq) \rangle = & B_{abc}(\bk,\bp,\bq)\, \delta(\bk+\bp+\bq)\\
& + \bar \phi_a \, P_{bc}(\bp) \,\delta(\bk) \delta(\bp + \bq) \, [\text{3 perm.}]\\
& + \bar \phi_a \bar \phi_b \bar \phi_c \, \delta(\bk) \delta(\bp) \delta(\bq)\, ,
\end{split}
\label{bisp} 
\ee 
and the trispectrum $Q_{abcd}(\bk,\bp,\bq,\br)$ by
\be
\begin{split}
\langle \phi_a(\bk) \phi_b (\bp) \phi_c(\bq) \phi_d(\br) \rangle =&   Q_{abcd}(\bk,\bp,\bq,\br)\, \delta(\bk+\bp+\bq+\br)\\
& + \bar \phi_a \, B_{bcd}(\bp, \bq, \br) \, \delta(\bk) \delta(\bp+\bq+\br)\, [\text{4 perm.}] \\
& + P_{ab}(\bk) P_{cd}(\bq) \, \delta(\bk + \bp) \delta(\bq+\br) \, [\text{3 perm.}] \\ 
& + \bar \phi_a \bar \phi_b \, P_{ab}(\bq) \, \delta(\bk) \delta(\bp) \delta(\bq+\br) \, [\text{6 perm.}] \\
& + \bar\phi_a \bar\phi_b \bar\phi_c \bar\phi_d \, \delta(\bk) \delta(\bp) \delta(\bq) \delta(\br)\, .
\end{split}
\label{trisp} 
\ee 
The brackets $\langle \ldots \rangle$ stand here for an averaging over an ensemble of events. The bispectrum, trispectrum etc. 
contain information about correlations between fluid dynamic fields at different space points (or wavevectors) at equal (Bjorken) time. 
In eqs.\ \eqref{bisp} and \eqref{trisp} we have not written out all permutations but indicated in square brackets how many there are of each kind.
We shall later neglect the effect of the trispectrum on the evolution of the spectrum and bispectrum. This is equivalent to setting
$Q_{abcd}(\bk,\bp,\bq,\br)=0$ in eq. (\ref{trisp}) and keeping only the disconnected parts of the four-point correlation function.

From the definitions it is clear that $P_{ab}(\bk) = P_{ba}(-\bk)$ whereas the bispectrum satisfies 
\be
B_{abc}(\bk,\bp,\bq)=B_{cab}(\bq,\bk,\bp)=B_{bca}(\bp,\bq,\bk),
\label{rot} \ee 
and similarly for the trispectrum. The $\delta$-functions in the definitions of the spectra are a consequence of our assumption of (statistical) homogeneity of space. 
Similarly, statistical isotropy implies that the
orientation of the vectors $\bk$, $\bp$, $\bq$ does not play any role. In particular we 
have 
\bea
P_{ab}(\bk)&=&P_{ab}(k)\, ,
  \nonumber \\
B_{abc}(\bk,\bp,\bq)|_{\bk+\bp+\bq=0} &=&B_{abc}(k,p,q).
\label{isotropy}  \eea

\subsection{Evolution equations for expectation values and spectra}
\label{sec3.3}

The evolution equations for the expectation value, spectrum, bispectrum and higher-order spectra result from the differentiation of 
eqs. (\ref{sp}), (\ref{bisp}) with respect to $t$. In principle this leads to an infinite hierarchy of equations, 
with the time evolution of the $n$-point correlation function involving the $(n+1)$-point function. 
This kind of hierarchy is analogous to the BBGKY hierarchy \cite{bbgky} 
and appears in related form also at other places in theoretical physics, 
for example in the context of the functional renormalization group \cite{erg}.

If one uses the definitions in eqs.\ \eqref{def:expvalue}-\eqref{trisp} and the evolution equation \eqref{eom} one finds for the expectation value
\begin{equation}
\partial_t {\bar\phi}_a = -\Omega_{ab}(0) \, \bar\phi_b+ \gamma_{abc}(0,0) \bar \phi_b \bar\phi_c
+ \int  \, d^2q \,\gamma_{abc}(-\bq,\bq)P_{bc} (\bq),
\label{eq:evexpvalue}
\end{equation}
for the spectrum
\begin{eqnarray}
\partial_{t}P_{ab}(\textbf{k})&=&-\tilde \Omega_{ac}(\bk) P_{cb}(\textbf{k})-\tilde\Omega_{bc}(-\bk) P_{ac}(\textbf{k})
\nonumber \\
&&+ \int d^{2}p \big[\gamma_{acd}(\bp,\bk-\bp)
B_{bcd}(-\textbf{k},\textbf{p}, \textbf{k}-\bp)
\nonumber \\
&&~~~~~~~~~~~+\gamma_{bcd}(\textbf{p},- \textbf{k}-\textbf{p})B_{acd}(\textbf{k},\textbf{p}, -\textbf{k}
-\textbf{p})\big] 
\, , \label{spectev1}
\end{eqnarray}
and for the bispectrum
\begin{eqnarray}
 \partial_{t}B_{abc}(\textbf{k},\textbf{p}, \textbf{q}){\big |}_{\bk+\bp+\bq=0}&=&
-\tilde\Omega_{ad}(\bk)B_{dbc}(\textbf{k},\textbf{p}, \textbf{q})-\tilde\Omega_{bd}(\bp)B_{adc}
(\textbf{k},\textbf{p}, \textbf{q})
\nn \\
&&-\tilde\Omega_{cd}(\bq) B_{abd}(\textbf{k},
\textbf{p}, \textbf{q})
\nonumber \\
&&
+2\big[\gamma_{ade}(-\textbf{p}, -\textbf{q}) P_{bd}(\textbf{p})P_{ce}(\bq)
+\gamma_{bde}(-\textbf{q}, -\textbf{k}) P_{cd}(\textbf{q})P_{ae}(\bk)
\nonumber \\
&&~~~~~
+\gamma_{cde}(-\textbf{k}, -\textbf{p}) P_{ad}(\textbf{k})P_{be}(\bp)
\big]\nonumber \\
&&+ \int d^2 r \big[ \gamma_{ade}(\br,\bk-\br) Q_{debc}(\br,\bk-\br,\bp,\bq) \nonumber\\
&&~~~~~~~~~~~ + \gamma_{bde}(\br,\bp-\br) Q_{deca}(\br,\bp-\br,\bq,\bk) \nonumber\\
&&~~~~~~~~~~~ + \gamma_{cde}(\br,\bq-\br) Q_{deab}(\br,\bq-\br,\bk,\bp) \big] \, .
\label{spectev2}
\end{eqnarray}
We have used in eqs.\ \eqref{spectev1} and \eqref{spectev2} the modified time evolution matrix 
\begin{equation}
\tilde \Omega_{ab}(\bk) = \Omega_{ab}(\bk) - \gamma_{acb}(0,\bk) \bar\phi_c - \gamma_{abc}(\bk,0) \bar\phi_c.
\end{equation}
For the particular vertex matrices defined by eq.~\eqref{gammas} and eqs.~\eqref{alpha1}-\eqref{beta} one finds that $\gamma_{abc}(0,\bk) = \gamma_{abc}(\bk,0) = 0$. This implies that $\tilde \Omega_{ab}(\bk) = \Omega_{ab}(\bk)$ in our case. Also in \eqref{eq:evexpvalue} we can drop the second term on the right hand side since $\gamma_{abc}(0,0) = 0$.\footnote{When inspecting eqs.~\eqref{alpha1}-\eqref{beta} in the limit $\bp\to 0, \bq\to 0$, one may worry about the fact that the value of $\beta(\bp,\bq)$ in eq.~\eqref{beta} depends on how this limit is taken. On the other hand, in eq.~\eqref{eq:evexpvalue}, $\gamma_{222}(\bp,\bq) = \beta(\bp,\bq)$ gives the effect of $\bar\phi_2^2$ on the time evolution of $\bar\phi_2$. We have argued in subsect.\ \ref{sec:defExpValuesSpectra} that one can always set $\bar\phi_2=0$ without affecting the physically relevant velocity field. Another possibility to understand this issue is to consider the velocity divergence $\theta=\partial_i v_i$ instead of the field $f=\phi_2$ as is discussed in subsect.\ \ref{sec:DerCouplingsUVProp}.}
The equations \ \eqref{eq:evexpvalue}, \eqref{spectev1}, \eqref{spectev2} could be written in slightly more compact form using eqs.\ \eqref{eq:realityconstraintOmegaGamma} and \eqref{eq:parityconstraintOmegaGamma}, but this is not needed for our discussion.

Note that the set of equations \eqref{eq:evexpvalue}-\eqref{spectev2} is not closed because the last equation depends on the trispectrum. The essential approximation that we make in the following in order to obtain a closed 
system is to neglect the effect of the trispectrum on the evolution of the bispectrum.

\subsection{Perturbative solution}

From now on, we assume $\tilde \Omega_{ab}(\bk) = \Omega_{ab}(\bk)$. As explained previously, this is realized 
explicitly in the model discussed here.  The formal solution of the above system is given by (we assume $\bar\phi_a (t_0)=0$)
\begin{eqnarray}
\bar\phi_a (t) &=& \int_{t_0}^t dt^\prime  \int d^2 q \, g_{ab}({\bf 0}\,,t,t^\prime) 
\gamma_{bcd}(-\bq,\bq)  P_{cd} (\bq,t'), \label{formalsol1e}
\\
 P_{ab}({\bf k}\,,t) &=& g_{ac}({\bf k}\,,t,t_0)  \, g_{bd}({\bf k}\,,t,t_0)  P_{cd}({\bf k}\,,t_0) \nonumber\\
&&  +\int_{t_0}^t dt^\prime  \int d^2 q \,g_{ae}({\bf k}\,,t,t^\prime) g_{bf}({\bf k}\,,t,t^\prime) \nonumber\\
&& \quad\times\left[  \gamma_{ecd}({\bf -q},\,{\bf q-k})\,B_{fcd}({\bf k},\,{\bf -q},\,{\bf q-k};\,t^\prime)\right.
\nonumber\\
&& \quad \quad
+ \left.\gamma_{fcd}({\bf -q},\,{\bf q-k})\,B_{ecd}({\bf k},\,{\bf -q},\,{\bf q-k};\,t^\prime)\right]\,,
\label{formalsol1} \\
B_{abc}({\bf k},\,{\bf -q},\,{\bf q-k};\,t)&=&
g_{ad}({\bf k}\,,t,t_0) \, g_{be}({\bf -q}\,,t,t_0) \,  g_{cf}({\bf q-k}\,,t,t_0)B_{def}({\bf k},\,{\bf -q},\,{\bf q-k};\,t_0)
\nonumber\\
&&+2 \int_{t_0}^t dt^\prime  \,g_{ad}({\bf k}\,,t,t^\prime) g_{be}({\bf -q}\,,t,t^\prime) g_{cf}({\bf q-k}\,,t,t^\prime)\nonumber\\
&&\quad \times\left[ \gamma_{dgh}({\bf -q},\,{\bf q-k})P_{eg}({\bf q}\,,t^\prime)P_{fh}({\bf q-k}\,,t^\prime)\right.\nonumber\\
&& \quad \quad+ \gamma_{egh}({\bf q-k},\,{\bf k})P_{fg}({\bf q-k}\,,t^\prime)P_{dh}({\bf k}\,,t^\prime)\nonumber\\
&& \quad \quad\left.
+ \gamma_{fgh}({\bf k},\,{\bf -q})P_{dg}({\bf k}\,,t^\prime)P_{eh}({\bf q}\,,t^\prime)
\right]\,,
\label{formalsol2}
\end{eqnarray}
where $g_{ab}({\bf k}\,,t,t^\prime)$ is the linear propagator, which gives the evolution of the field at the linear level: $\phi^L_a({\bf k}, t) = g_{ab}({\bf k}\,,t,t^\prime) \phi^L_b({\bf k}, t^\prime)$. According to the definition (\ref{eq:deft}), we have $t_0=3\tau_0/4$.
The propagator satisfies
\be
\partial_t g_{ab}({\bf k}\,,t,t^\prime) = -\Omega_{ac}(\bk,t) g_{cb}({\bf k}\,,t,t^\prime), 
~~~~~~~~~~~~~ g_{ab}({\bf k}\,,t,t)=\delta_{ab}.
\label{propprop} \ee

The solutions can be expanded in powers of the interaction vertex $\gamma_{abc}$ in order to 
establish the connection with perturbation theory.
The lowest order, corresponding to linear theory, is obtained by setting $\gamma_{abc}=0$. The linear spectrum $ P^L_{ab}$ 
and bispectrum $ B^L_{abc}$ 
are given by the first line of each of the above equations. 
The $O(\gamma)$ correction to the bispectrum is obtained by inserting $ P^L_{ab}$ in place of $P_{ab}$ in the r.h.s. of eq.~(\ref{formalsol2}). Inserting the bispectrum at this order in eq.~(\ref{formalsol1}) generates the $O(\gamma)$ and $O(\gamma^2)$ contributions to the power spectrum. 
Iterating the procedure generates the higher-order corrections. However, differences with perturbation theory
arise at higher orders, because of  the approximation $Q_{abcd}=0$ that we have made in deriving the evolution equations 
for the power spectrum and bispectrum.

\subsection{Linear propagator}
\label{subsec:LinearPropagator}

In order to obtain an analytical expression for the propagator, which is necessary for the iterative solution of eqs. 
(\ref{formalsol1}), (\ref{formalsol2}), we consider 
eqs. (\ref{fouf}), (\ref{foud}), neglecting the nonlinear terms. They can be put in the form
\bea
\dot{\td}_{\bk}&=&-\frac{1}{2t} \td_{\bk}+  \frac{A_k}{\sqrt{t}}   \tf_{\bk}
\label{foufl}\\
\dot{\tf}_{\bk}&=&-\td_\bk-B_k \tf_\bk,
\label{foudl}
\eea
with $A_k= \left({\tau_0}/{12 } \right)^{1/2}    k^2$,
$B_k=4 \nu_0 k^2/3$.
Through the definition $\delta_\bk=\sqrt{t}\td_\bk$, eq. (\ref{foufl}) becomes 
\be
\dot{\delta}_\bk=A_k \tf_\bk\, ,
\ee
while
eq. (\ref{foudl}) gives 
\be
\ddot{\delta}_\bk+B_k\dot{\delta}_\bk+\frac{A_k}{\sqrt{t}}\delta_\bk=0.
\label{foudln} \ee
Finally, the definition $\chi_\bk=\exp(B_kt/2)\, \delta_\bk$ leads to
\be
\ddot{\chi}_\bk+\left(\frac{A_k}{\sqrt{t}}-\frac{B_k^2}{4}\right) \chi_\bk=0.
\label{foudlnf} \ee

This equation has simple analytical solutions when either one of the two terms in the parenthesis dominates. 
The dominance switches from the 
first to the second term at a time
$\tau/\tau_0\simeq (4\nu_0^2 k^2/3)^{-3/2}.$ Approximating the kinematic viscosity by its homogeneous value (\ref{eq:rescaling}), and
assuming an equation of state $\epsilon=3p$, we find from eq. (\ref{eq:Bjorkennutau}) that 
\be
{\nu_0^2}{k^2}\simeq \left(\frac{\eta}{s} \right)^2 \frac{k^2}{T^2_\text{Bj}(\tau_0)}.
\label{esti} \ee
We expect that $\tau_0 \simeq 1$ fm, $T_\text{Bj}(\tau_0) \sim 200-500$ MeV, while the relevant momenta satisfy $k \lta$ 1 fm$^{-1}$. 
For values of $\eta/s \sim 1/(4\pi)$ the first term in (\ref{foudln})
dominates for the times of interest, which are $t\lta 10$ fm. In the limit that the second term is neglected ($B_k=0$), 
the solution of eq. (\ref{foudlnf}) is
\be
\chi_{\bk}(t)=\sqrt{t} \left[ c_1 J_{-\frac{2}{3}}\left(\frac{4}{3} \sqrt{A_k} t^{3/4}\right)
+c_2  J_{\frac{2}{3}}\left(\frac{4}{3} \sqrt{A_k} t^{3/4}\right)\right],
\ee
(where $J_n(x)$ are the Bessel functions of the first kind) so that 
\be
\td_{\bk}(t)=\exp\left(-\frac{1}{2}B_kt \right) \left[ c_1 J_{-\frac{2}{3}}\left(\frac{4}{3} \sqrt{A_k} t^{3/4}\right)
+c_2  J_{\frac{2}{3}}\left(\frac{4}{3} \sqrt{A_k} t^{3/4}\right)\right]
\label{dbk} \ee
and 
\be
\tf_\bk(t)=\frac{\sqrt{t}}{A_k}\left(\dot{\td}_\bk+\frac{\td_\bk}{2t}\right).
\label{fdk} \ee
These expressions indicate oscillatory behavior corresponding to propagating sound waves in the plasma.
The exponential suppression at late times describes the attenuation of the high-frequency modes through the effect of viscosity.

\begin{figure}[t]
\begin{minipage}{72mm}
\includegraphics[width=\linewidth,height=50mm]{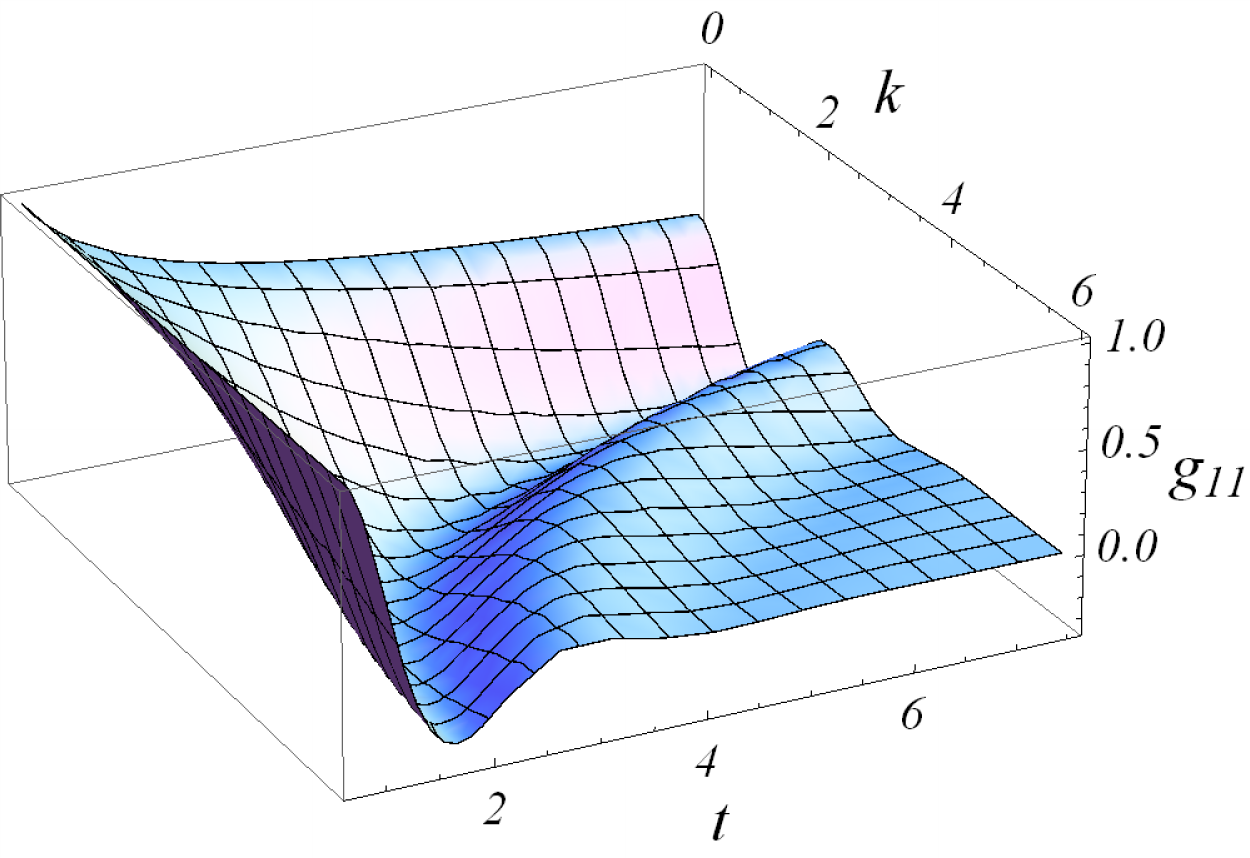}
\caption{
{The analytical approximation for the 11-component of the perturbative propagator
$g_{11}(k,t,t_0)$, 
evaluated through the solution (\ref{dbk}), (\ref{fdk}) with $t_0=3\tau_0/4$, $\tau_0=1$ fm and $\nu_0=0.04$ fm. (All quantities measured in fm.)}}
\label{prop11}
\end{minipage}
\hfil
\begin{minipage}{72mm}
\includegraphics[width=\linewidth,height=50mm]{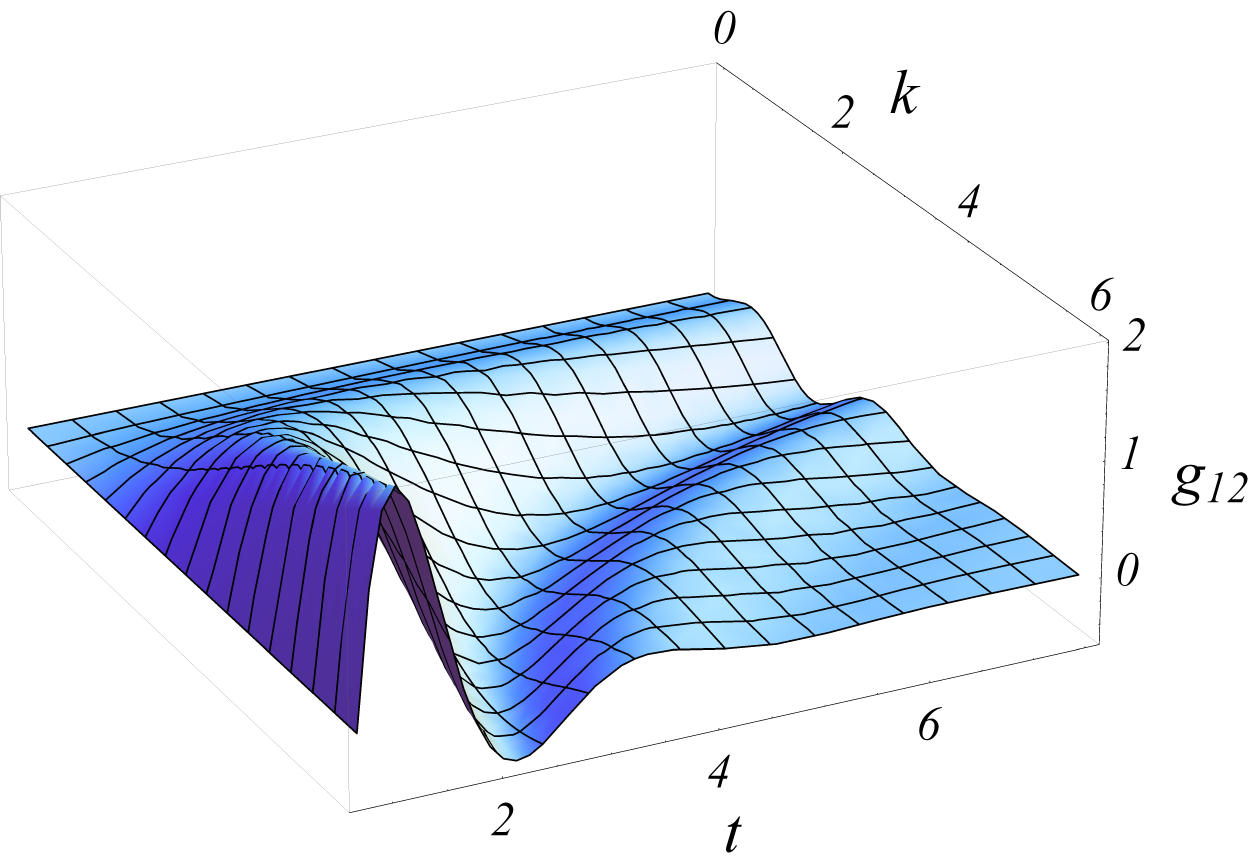}
\caption{
{The analytical approximation for the 12-component of the perturbative propagator
$g_{12}(k,t,t_0)$, 
evaluated through the solution (\ref{dbk}), (\ref{fdk}) with $t_0=3\tau_0/4$,
$\tau_0=1$ fm and $\nu_0=0.04$ fm. (All quantities measured in fm.)}}
\label{prop12}
\end{minipage}
\end{figure}

The propagator can be determined easily by fixing the constants $c_1$, $c_2$ through the initial conditions
$\td_\bk(t_0)$, $\tf_\bk(t_0)$. The four components of $g_{ab}$ are given by the coefficients of 
$\td_\bk(t_0)$, $\tf_\bk(t_0)$ in the resulting expressions. We do not display the explicit form of the propagator because of its 
length and complicated structure.
We have verified numerically that eqs. (\ref{dbk}), (\ref{fdk}) provide a very good approximation for all
scales and times relevant for our analysis. 

In figs. \ref{prop11} and \ref{prop12} we display the 11- and 12-components of the 
propagator for $\tau_0=1$ fm and $\nu_0=0.04$ fm. These figures document the appearance of waves and the attenuation 
of the high-frequency modes.

The explicit analytical expressions derived above allow one to deduce certain quantitative features of the solutions, which are also
reflected in the evolution of the spectra. The Bessel functions $J_{\pm\frac{2}{3}}(z)$ display oscillatory behavior for real $z$.
This corresponds to the appearance of acoustic waves in the relativistic fluid, supported by the strong pressure $p=\epsilon/3$. 
The wavelength of the oscillations has a time dependence that can be extracted from eq. (\ref{dbk}). 
Its scale is set by the first zero of the Bessel functions. For $J_{-\frac{2}{3}}(z)$ this is located at $z_{1-}=1.24$, and
for $J_{\frac{2}{3}}(z)$ at $z_{1+}=3.38$. The valleys and ridges apparent in figs. \ref{prop11}, \ref{prop12} 
occur at values of order 1 for the combination $k\, t^{3/4}$ that appears in the argument of the Bessel functions in eq. (\ref{dbk}).
Another feature is the attenuation of the high-frequency modes induced by the viscosity. The characteristic time for this effect
is set by the exponential in eq. (\ref{dbk}). As $B_k\sim \nu_0k^2$, the high-frequency modes quickly die out within a time
$\sim 1/(\nu_0 k^2)$. This feature is also apparent in figs. \ref{prop11}, \ref{prop12}.

\section{Initial conditions}
\label{sec:InitialConditions}

Before turning to the discussion of solutions of the equations of motion \eqref{eq:evexpvalue}, \eqref{spectev1}, \eqref{spectev2}
for the expectation value, spectrum and bispectrum, we introduce in this section the initial conditions from which these equations
of motions are evolved.

\subsection{Modelling the initial fluctuations}

The evolution equations are initialized at the time $\tau_0$ at which a fluid description becomes approximately valid. The initial values 
at $\tau_0$ are a result of initial-state physics and of the early non-equilibrium dynamics that drives (local) thermalization. They are 
to be determined from a theoretical description of that era. Here, we obtain them from a simple model inspired by a Glauber-type 
description. 
As in most of the phenomenological literature, 
we assume that only the density fluctuates initially and we neglect possible initial flow fluctuations. 
To model the density $d$ of transverse fluctuations, 
we assume that the deviations from the Bjorken background are small so that 
\begin{equation}
d = \ln \left( \frac{T}{T_\text{Bj}} \right) \approx  \frac{w - w_\text{Bj}}{4 w_\text{Bj}}\, ,
\label{Bjbkg}
\end{equation}
where we have assumed a velocity of sound $c_S^2 = 1/3$. 

The transverse enthalpy density $w$ is assumed to originate from a number of localized sources at positions that we take to be 
independently and uniformly distributed in the transverse plane. For a single event, the corresponding density field $d$ is given 
by~\cite{Yan:2013laa,Floerchinger:2014fta}
\begin{equation}
d(\bx) = -\frac{1}{4} + \frac{A}{4  N}\sum_{j=1}^N f(\bx-\bx_j).
\end{equation}
Here, the index $j$ labels different sources or scattering centers, and the strength of the source is given by a smoothened 
transverse density $f(\bx - \bx_j)$ smeared around $\bx_j$. In the following, we consider the case that all sources
have the same normalized strength, $\int d^2x\, f(\bx) = 1$. We note that in a Glauber type description, one may associate each source to a nucleon participating in
inelastic processes, or to a pair of participants. In this work, we keep the number of sources $N$ fixed. But the model could 
be easily extended to include fluctuations in the number of contributing sources or in the strength (i.e. norm) of each source. 
In that case, it would account for fluctuations in multiplicity. In agreement with the symmetries of the Bjorken model, 
we take all sources to be uniformly distributed in a transverse area $A$ which corresponds to a probability distribution $p(\bx)=1/A$.
The expectation value of the transverse density is independent of the transverse position. The Bjorken background entering (\ref{Bjbkg})
can be chosen such that by construction
\begin{equation}
\langle d(\bx) \rangle = 0\, .
\end{equation}

\subsection{The initial spectra}

To calculate correlation functions, we introduce the partition function~\cite{Floerchinger:2014fta}
\begin{eqnarray}
Z[J]  &=& \left\langle e^{\int d^2x J(\bx) d(\bx)} \right\rangle = e^{-\tfrac{1}{4} \int d^2x J(\bx)} \left\langle \prod_{j=1}^N e^{\frac{A}{4 N}  \int d^2x \, J(\bx) f(\bx- \bx_j)} \right\rangle 
	\nonumber \\
	&=& e^{-\frac{1}{4} \int d^2x J(\bx)} \left[ \frac{1}{A}\int d^2 x^\prime \,  e^{\tfrac{A}{4 N}\int d^2 x\, J(\bx) f(\bx-\bx^\prime)} \right]^N
        \nonumber \\
 	&=& e^{-\tfrac{1}{4} (2\pi)^2 \tilde J(0)} \left[ \frac{1}{A} \int d^2 x \; e^{\frac{A}{4 N}(2\pi)^2
	\int d^2 k  \tilde J(-\bk) \tilde f(\bk) e^{-i \bk \bx}} \right]^N.
\end{eqnarray}
To simplify this partition function, we have made use of the assumption that the random variables $\bx_j$ are independent and that they are uniformly distributed.
We have introduced the Fourier transforms $\tilde J(\bk)$ and $\tilde f(\bx)$ of the source $J(\bx)$ and the density $f(\bx)$ of an individual collision center, respectively.

Correlation functions can now be calculated as usual, by taking functional derivatives. For example, we find 
\begin{equation}
\begin{split}
\langle \tilde d(\bk) \tilde d(\bk^\prime) \rangle & = \frac{1}{(2\pi)^4} 
\frac{\delta^2}{\delta \tilde J(-\bk) \delta \tilde J(-\bk^\prime)} \ln Z[J] {\big |}_{J=0}\\
& =  (2\pi)^2 \frac{A}{16 N} \left[  |\tilde f(\bk)|^2 - \tilde f(0)^2 \frac{(2\pi)^2 \delta^{(2)}(\bk)}{A}\right]\delta^{(2)} (\bk+ \bk^\prime).
\end{split}
\label{eq:twopointfunct}
\end{equation}
Note that although the second term in the last line seems to be singular for $\bk=0$, it is actually finite, because 
one has $(2\pi)^2 \delta^{(2)}(\bk) / A \to 1$. 
For the particular case of the Gaussian source for which
\be
\tilde{f}(\bk)=\frac{1}{(2\pi)^2}e^{-\frac{\sigma^2 k^2}{2}},
\label{fourier} \ee
one finds 
the initial spectrum of density fluctuations
\begin{equation}
\langle \tilde d(\bk) \tilde d(\bk^\prime) \rangle = P_d(k)\, \delta^{(2)}(\bk+\bk^\prime),
\end{equation}
with
\begin{equation}
\begin{split}
P_d(k) & = \frac{1}{(2\pi)^2} \frac{A }{16N} \; e^{-\sigma^2 k^2} \quad\quad\quad  {\rm for}\,\,k>0 \\
P_d(0) & = 0.
\end{split}
\label{eq:InitialSpectrum}
\end{equation}
In the limit of point-like sources ($\sigma\to 0$) the initial spectrum 
becomes independent of $k$. It is clear that this limit is somewhat artificial. 
In reality one expects that the enthalpy due to each scattering center is smooth and non-singular. This is a consequence of 
the finite size of the nucleon itself, but also of the early dynamics that drives local thermalization. 
Soon after the collision the dissipative effects are expected to be large, which leads to a fast smoothening of the enthalpy density. 
If a fluid dynamic description becomes valid at time $\tau_0$, one expects that spatial structures on much finer scales than $c\tau_0$ have already disappeared. (For concreteness, one might think of dynamics close to free streaming.) 
Based on these considerations, one expects that $\sigma\approx \tau_0$ is reasonable.

Higher-order correlation functions can be calculated in analogy to eq.\ \eqref{eq:twopointfunct}. For example, the three-point function is given by
\begin{equation}
\langle \tilde d(\bk_1) \tilde d(\bk_2) \tilde d(\bk_3) \rangle = B_d(k_1, k_2, k_3) \, \delta^{(2)}(\bk_1 + \bk_2 + \bk_3)
\end{equation}
with
\begin{equation}
\begin{split}
 B_d(k_1, k_2, k_3) & = (2\pi)^2 \frac{A^2}{64 N^2}  {\Bigg [} \tilde f(\bk_1) \tilde f(\bk_2) \tilde f(\bk_3) \\
&  - \tilde  f(\bk_1) \tilde f(\bk_2) \tilde f(0) \frac{(2\pi)^2 \delta^{(2)}(\bk_3)}{A} \; [3\; \text{perm.}]  
+ 2 \tilde f^3(0) \frac{(2\pi)^4 \delta^{(2)}(\bk_1) \delta^{(2)}(\bk_2)}{A^2} {\Bigg ]}
\end{split}
\label{thrrep} \ee
In the second line we have written explicitly only one of three permutations of $\bk_1$,$\bk_2$, $\bk_3$.
For nonzero momenta and the Gaussian source we have 
\be
 B_d(k_1, k_2, k_3)=\frac{1}{(2\pi)^4} \frac{A^2}{64 N^2} \; e^{-\frac{\sigma^2}{2} \left( k_1^2 + k_2^2 +k_3^2 \right)}.
\label{threepg}
\end{equation}

\section{UV sensitivity and the range of applicability of fluid dynamics}
\label{sec5}

Fluid dynamics is the long wavelength limit of a consistent transport theory, and thus it needs not 
be well-behaved in the UV. Understanding the sensitivity of the fluid dynamic evolution on UV cut-offs is therefore important for determining its range of 
applicability. In the model studied here, the nonlinear couplings $\gamma_{abc}(\bp,\bq)$ that enter the equations of motion \eqref{eq:evexpvalue}, 
\eqref{spectev1}, \eqref{spectev2} arise from the spatial-derivative terms in the fluid dynamic equations. They show power-law growth in the UV. 
In principle, there are two physical mechanisms that result in UV cutoffs. First, density fluctuations shortly 
after the collision are expected to be limited to wavenumbers of order $|\bk|\lsim 1 /\text{fm}$ and 
the initial conditions are therefore UV-tame on physical grounds. Second, dissipative processes are expected 
to provide for a dynamic mechanism that further damps high-momentum modes. However, despite these two UV-regulating mechanisms, one cannot exclude 
a relatively strong sensitivity of the solutions on the details of the initial conditions in the UV. 
Therefore, before  turning in the next section to the numerical study of the equations of motion \eqref{eq:evexpvalue}, \eqref{spectev1}, \eqref{spectev2},
we discuss here the UV behavior of the fluid dynamics for density perturbations, as introduced in section \ref{correl}. 

\subsection{Derivative couplings and UV properties}
\label{sec:DerCouplingsUVProp}

We emphasize first that the UV sensitivity of the fluid dynamic evolution cannot be attributed solely to nonlinear terms in the iterative solution of 
eqs. (\ref{formalsol1})-(\ref{formalsol2}). This becomes apparent if one rewrites the equation of motion (\ref{eom})
for $\phi_a(t,\bk) = \left(\tilde d_\bk, \tilde f_\bk \right)$ into an equation for the fluid dynamic fields
$\tilde \phi_a(t,\bk) = \left(\tilde d_\bk, \tilde \theta_\bk \right)$, where the divergence of the velocity field $\theta$ 
with $\tth_\bk=-k^2 \tf_\bk$ replaces the field $f$. (An analogous formulation is used for the analysis of cosmological 
perturbations in the formalism that we are employing \cite{time}.) The equations of motion for $\tilde \phi_a(t,\bk)$
are of the same form as (\ref{eom}), but with couplings defined by 
\bea
\tilde{\alpha}_1(\bp,\bq)&=&-\frac{\bp \bq}{q^2} \, ,
\label{alpha1t} \\
\tilde{\alpha}_2(\bp,\bq)&=&\frac{2 \nu_0}{3} \left( \frac{\left(\bp \bq \right)^2}{p^2 q^2}-\frac{1}{3}\right) \, ,
\label{alpha2t} \\ 
 \tilde{\beta}(\bp,\bq)&=&\frac{1}{p^2 q^2}\left[ \left(\bp\bq \right)^2-\frac{1}{3}p^2q^2+\frac{1}{3} \left(\bp \bq \right) \left(q^2+p^2\right)
\right]\, .
\label{betat} 
\eea 
These couplings are dimensionless. For large momenta, they are much better behaved than (\ref{alpha1}) - (\ref{beta}), although 
there are still divergent directions in momentum space. (For example, the coupling (\ref{alpha1t}) diverges for large $\bp$ and fixed $\bq$.) 
On the other hand, the linearlized evolution is now described by the equations
\bea
\dot{\td}_{\bk}&=&-\frac{1}{2t} \td_{\bk}- \sqrt{\frac{\tau_0}{12t}}   \tth_{\bk}\, ,
\label{fouflt}\\
\dot{\tth}_{\bk}&=&k^2 \td_\bk- \frac{4 \nu_0}{3} k^2 \tth_\bk\, .
\label{foudlt}
\eea
This illustrates that it depends on the choice of independent fluid dynamic fields whether the strong momentum dependence 
is found in the linear or nonlinear terms of the evolution equation. In eq. (\ref{eom}) for $\phi_a$, the nonlinear couplings grow
like power-laws with increasing momenta. In eq.~(\ref{foudlt}), this strong UV-dependence has been reshuffled now mainly into 
the linear evoluton of $\dot{\tth}_{\bk}$,
The same equation (\ref{foudlt}) also illustrates clearly that the reason for this strong momentum dependence is physical. 
The first term on its r.h.s. indicates that  strong flows are induced by density inhomogeneities with large momenta. 
This is of course a very physical effect: the fluid is accelerated by pressure gradients. The second term results in an exponential suppression
of the high-momentum fluctuations through the viscosity. The general evolution involves a competition between these two terms. 

It is easy to verify that the use of $\td_\bk$, $\tth_\bk$, instead of $\td_\bk$, $\tilde{f}_\bk$, 
leads to identical results. For our numerical solution we employ the fields
$\td_\bk$, $\tf_\bk$, mainly because of the symmetry of the effective couplings (\ref{alpha1})-(\ref{beta}) under the 
exchange $\bp \leftrightarrow \bq$. 

\subsection{Mode-mode coupling and the influence of UV modes}
To obtain a better understanding of the UV sensitivity of fluid dynamic perturbation theory, 
we consider here a particularly simple example where some analytic insight is possible, namely
the nonlinear corrections to the spectrum $P_{ab}(k,t)$ with $k\to 0$.
We concentrate on the  $O(\gamma)$ correction to  the density-density
component $P_{11}(0,t)$ in the presence of an initial bispectrum
whose only non-vanishing component is $B_{111}$. The spectrum can be expressed as
\be
P_{11}(0,t)=g_{1c}(0,t,t_0)g_{1d}(0,t,t_0) P_{cd}(0,t_0)+
\frac{2 \pi}{t} \int_{t_0}^t dt'\int_0^{\infty} dq \, h(q,t',t_0)B_{111}(0,-\bq,\bq)
\label{p11est} \ee
with 
\be
h(q,t',t_0)=\sqrt{t_0 t'}\left(-4q^3  \, g_{11}(q,t',t_0) \, g_{21}(q,t',t_0) 
+  \frac{16}{9}q^5 \nu_0  \, g_{21}(q,t',t_0)^2 
\right).
\label{funh} \ee
For $\bk=0$ the propagator simplifies to
\be
g(0,t,t')= \left( \begin{array}{cc}
\sqrt{t'/t}  &\hspace{0.5cm} 0  \\
-2 \sqrt{t't}+2t' & \hspace{0.5cm} 1   \end{array} \right)\, ,
\ee  
and the couplings are 
$\gamma_{112}(-\bq,\bq)=-q^2/2$ and 
$\gamma_{122}(-\bq,\bq)=4 q^4 \nu_0/9$.

\begin{figure}[!h]
\centering
\includegraphics[width=72mm,height=50mm]{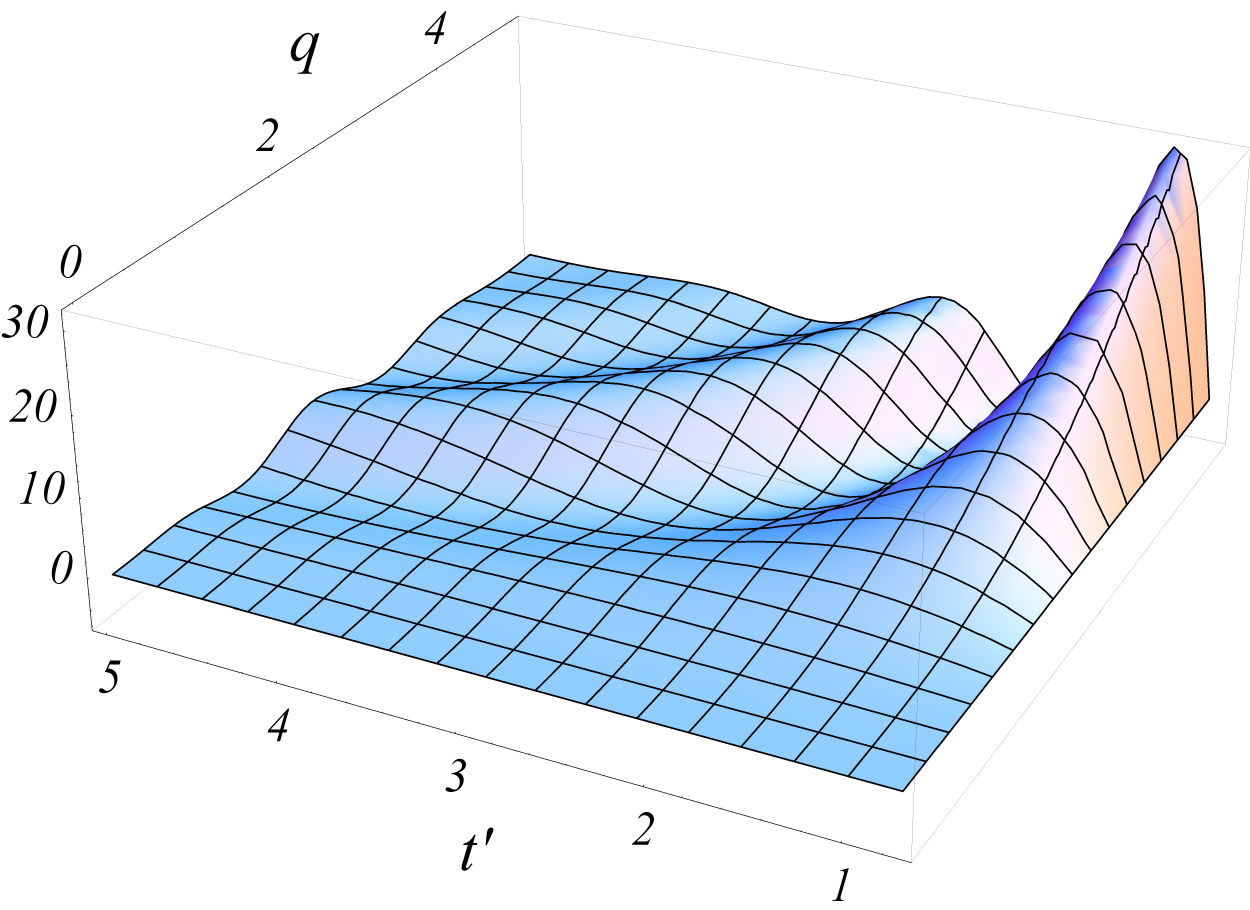}
\caption{
{$h(q,t',t_0)$ for $t_0=3\tau_0/4$, $\tau_0=1$ fm, $\nu_0=0.04$ fm. For better visibility, the momentum and time axes have been exchanged compared to figs. \ref{prop11}, \ref{prop12}. (All quantities measured in fm.)}}
\label{h}
\end{figure}

In general, nonlinear corrections in the evolution of the spectra arise from the coupling between modes with different momenta. 
This coupling also results in the transfer of energy between modes. Since much of the energy is initially concentrated 
in the high-momentum modes, one expects that the nonlinear evolution will induce the enhancement of the low-momentum ones.
We now turn to the numerical solution of (\ref{p11est}) in order to illustrate these expectations. Numerical results
will be presented for $t_0=3/4$ fm (corresponding to $\tau_0=1$ fm) and $\nu_0=0.04$ fm, and for variations around these
default values, where indicated. 

\begin{figure}[h]
\begin{minipage}{72mm}
\includegraphics[width=\linewidth,height=50mm]{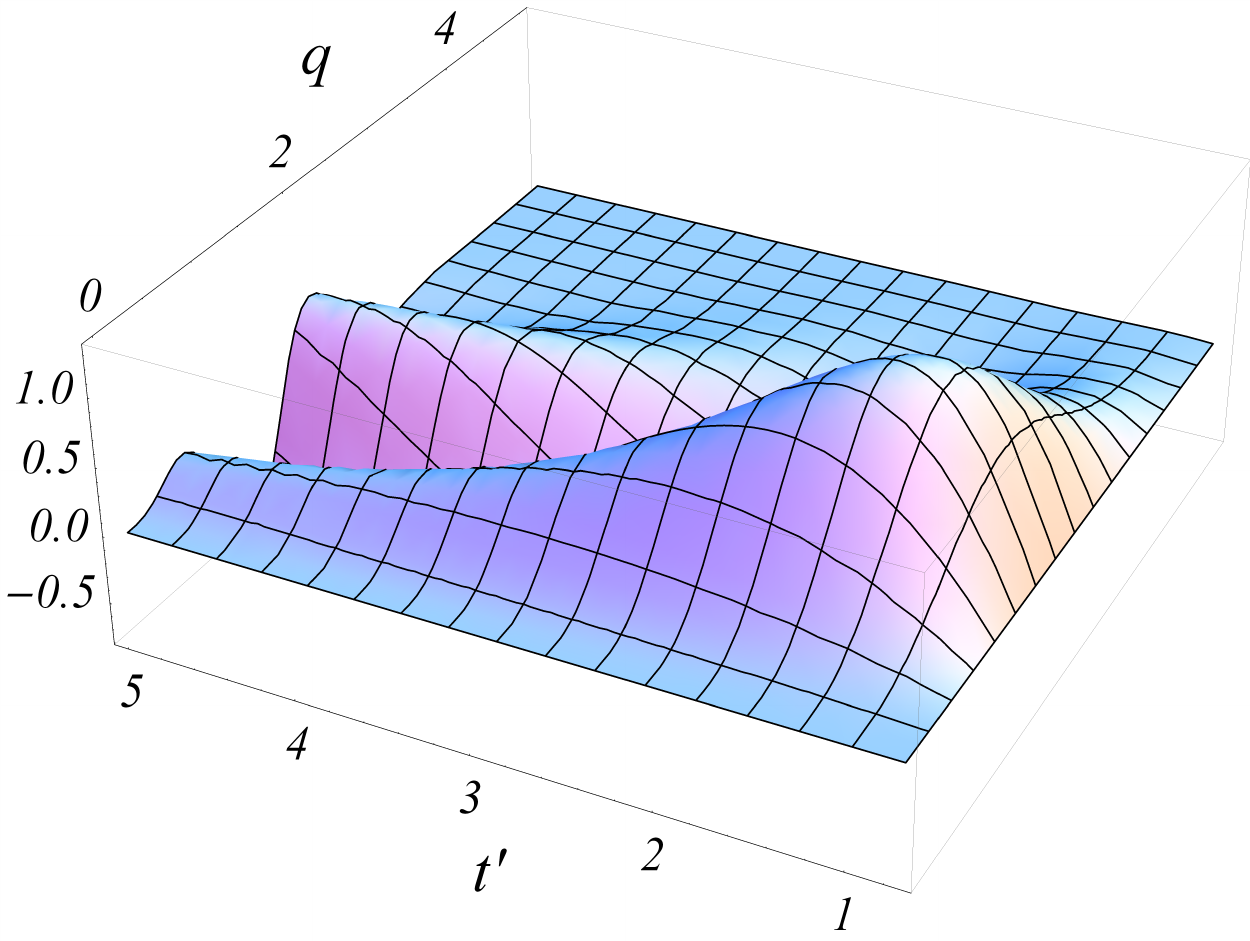}
\caption{
{$h(q,t',t_0)\exp(-\sigma^2 q^2)$ for $t_0=3\tau_0/4$, $\tau_0=1$ fm, $\nu_0=0.04$ fm, $\sigma=0.4$ fm. (All quantities measured in fm.)}}
\label{hB04}
\end{minipage}
\hfil
\begin{minipage}{72mm}
\includegraphics[width=\linewidth,height=50mm]{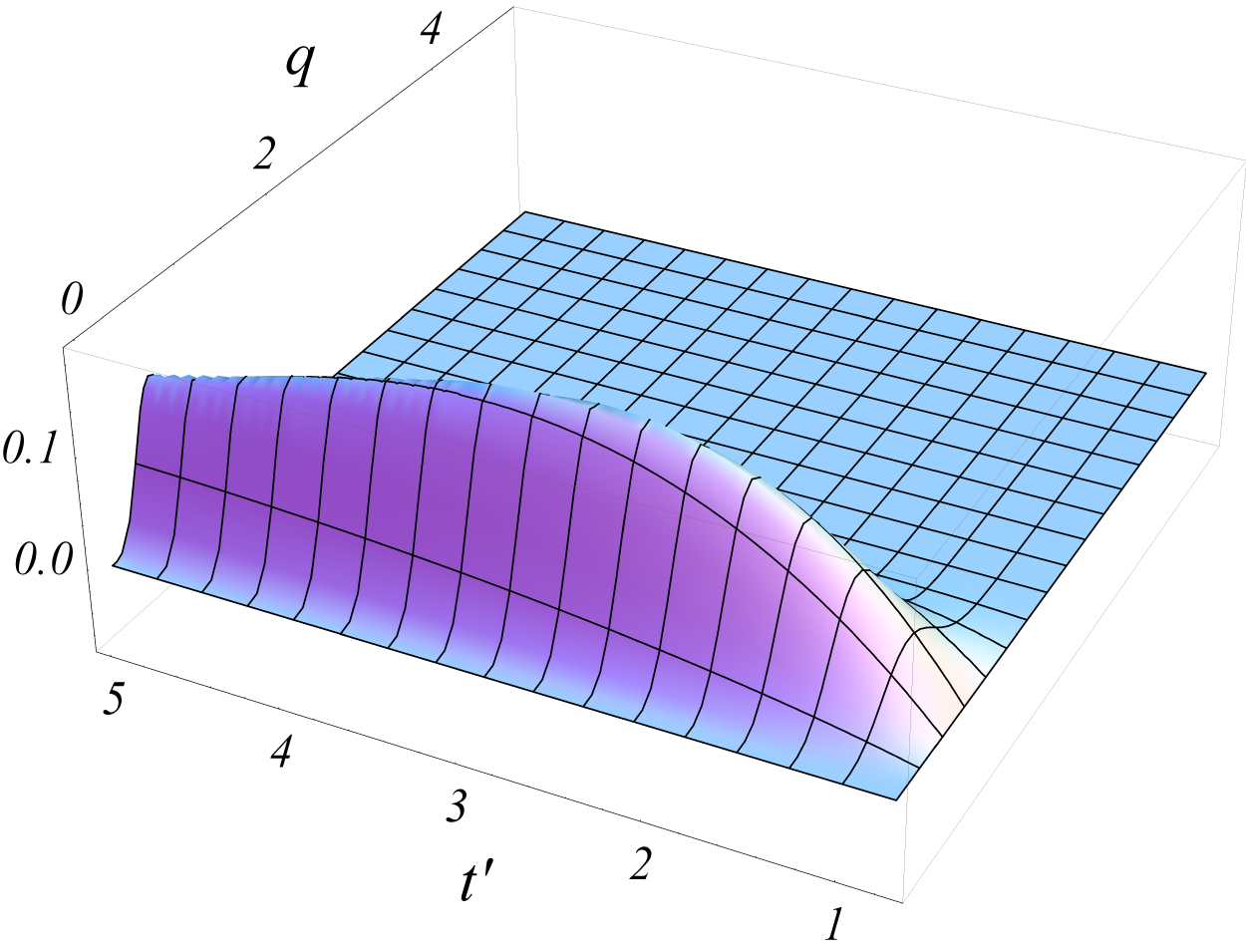}
\caption{
{$h(q,t',t_0)\exp(-\sigma^2 q^2)$ for $t_0=3\tau_0/4$, $\tau_0=1$ fm, $\nu_0=0.04$ fm, $\sigma=1$ fm. (All quantities measured in fm.)}}
\label{hB1}
\end{minipage}
\end{figure}

In fig. \ref{h}, we plot the function $h(q,t',t_0)$ of eq.~(\ref{funh})
that determines the size of nonlinear corrections to the spectrum $P_{11}(0,t)$. 
At early times, $h(q,t',t_0)$ is seen to grow very fast with increasing momenta $q$. At later times, dissipative effects due to the finite shear
viscosity dampen these high-momentum modes rapidly. However, despite this viscous damping, the momentum integration in the second 
term of eq. (\ref{p11est}) would lead to a divergent result for a constant initial bispectrum $B_{111}(0,-\bq,\bq)$. To see this, we note
first that as a consequence of the exponential suppression factor in eq. (\ref{dbk}), the propagator $g_{ab}(q,t',t_0)$ that defines $h(q,t',t_0)$ takes 
very small values when $\nu_0 q^2 t' \gta 1$. For large $q$ and constant $B_{111}(0,-\bq,\bq)$, the leading contribution to the integral of eq. (\ref{p11est}) 
comes from the thin region with $t'-t_0 \lta 1/(\nu_0 q^2)$. On the other hand, the two terms in the function $h(q,t',t_0)$ involve powers of $q$ that 
overcompensate the suppression by the  $q^2$, so that the integral is UV divergent. The spectra $P_{12}$ and $P_{22}$ display similar behavior. 
This illustrates that UV-tame initial conditions are a prerequisite for the validity of hydrodynamic perturbation theory. The physically motivated
initial conditions discussed in section ~\ref{sec:InitialConditions} satisfy this requirement. In particular, an initial bispectrum of the form (\ref{threepg}) 
amounts to an exponential suppression of the large-$q$ modes. In figs. (\ref{hB04}) and (\ref{hB1}) we illustrate that the resulting integrand 
$h(q,t',t_0) \exp(-\sigma^2 q^2)$ entering the $q$-integration in (\ref{p11est}) vanishes now at large $q$. The scale at which this damping
occurs is determined by the physics defining initial conditions (which is parametrized by $\sigma=0.4$ and 1 fm, respectively), but for the 
entire class of initial conditions, the result is UV finite.

The above discussion clarifies how the general expectation that the applicability of fluid dynamics is limited to sufficienlty small momentum modes
is realized technically for the particular case of evolving $P_{11}(0,t)$. For the general case, however, analytical insight is more difficult
to obtain, and it is difficult to check {\itshape a priori} the convergence of the iterative solution of eqs. (\ref{formalsol1}), (\ref{formalsol2}). The numerical analysis can then
provide a consistency check {\itshape a posteriori}. 

\subsection{Range of applicability of fluid dynamics}

In the previous subsection, we emphasized that fluid dynamics can apply only if initial conditions are 
regulated in the UV. We now turn to estimating the scale below which this UV cutoff must lie for initiating
a consistent fluid dynamic evolution. To this end, we consider corrections
to the relativistic Navier-Stokes equations. In a second-order formalism, the
constitutive relation for the shear stress
\begin{equation}
\pi^{\mu\nu} +2\, \eta \, P^{\mu\nu}_{\;\;\;\;\alpha\beta} \nabla^\alpha u^\beta =0
\end{equation}
($P^{\mu\nu}_{\;\;\;\;\alpha\beta}$ is the standard projector to the symmetric,
transverse and traceless part of a tensor) gets supplemented by additional terms, in
particular by
\begin{equation}
\tau_\pi \left( P^{\mu\nu}_{\;\;\;\;\alpha\beta} \, u^\gamma \nabla_\gamma
\pi^{\alpha\beta} + \frac{4}{3} \pi^{\mu\nu} \nabla_\gamma u^\gamma \right).
\end{equation}
Within the approximations made in section 2 one finds that the relaxation term becomes important when
\begin{equation}
\tau_\pi | \dot{\tilde f}_\bk | \gsim |\tilde f_\bk |.
\label{eq:relaxationtimeterm}
\end{equation}
As a rough estimate, one may assume that the time dependence of $\tilde f_\bk$ is
dominated by sound propagation with frequency $\omega = c_S |\bk| $. Eq.\ \eqref{eq:relaxationtimeterm} corresponds then to
$\tau_\pi c_S |\bk| \gsim 1$, and the range of applicability of fluid dynamics can be expected to cover momenta
\begin{equation}
 |\bk| \lsim \frac{1}{\tau_\pi c_S}\, .
\end{equation}
In general, both the relaxation time $\tau_\pi$ and the velocity of sound $c_S$ will depend on the microscopic 
dynamics of the f luid. For the sound velocity, however, this dependence is relatively mild, and we assume the
value $c_S = 1/\sqrt{3}$ for an ideal equation of state for the purpose of the following estimates. The relaxation time $\tau_\pi$
is not known from first principles in QCD, but it is known for several non-abelian gauge theories
with gravity duals in both the strong and weak coupling limit. It is expected to behave similarly in QCD.
In particular, for ${\cal N}=4$ SYM theory, it is $\lim_{\lambda \to \infty} \tau_\pi = (2 -\ln 2)/(2\pi T)$ in the limit
of infinite t'Hooft coupling $\lambda = g^2 N_c$ ~\cite{Baier:2007ix}.
Assuming this value, i.e. assuming a strongly coupled liquid, we find 
\begin{equation}
|\bk| \lsim \frac{2 \pi \sqrt{3}}{2-\ln 2} T \approx 8.3 \,  T.
\end{equation}
A similar numerical estimate is obtained, if one assumes that the time dependence of $\tilde f_\bk$ is dominated by the viscous
damping rate $4\nu k^2/3$, where $\nu=\eta/(sT)$ is the kinematic shear viscosity. In this case, the second order term remains
small for 
\begin{equation}
\tau_\pi \frac{4\eta}{3 sT} k^2 \lsim 1,
\end{equation}
and using $\eta/s = 1/(4\pi)$, one finds $|\bk| \lsim \sqrt{\frac{6\pi^2}{2-\ln 2}} \approx 6.7 \,T$ for the range of applicability of
fluid dynamics. 

It is interesting to note that the strong coupling limit of $\tau_\pi$ used above can be written in the form
$\lim_{\lambda \to \infty} \tau_\pi \simeq (2.7/T) (\eta/s)$, (where $\lim_{\lambda \to \infty} (\eta/s) = 1/4\pi$)
while the weak coupling limit is known to satisfy $\lim_{\lambda \to 0} \tau_\pi \simeq (5.9/T) (\eta/s)$~\cite{York:2008rr}. Such a monotonous
relation between relaxation time and sound attenuation length is also expected on physics grounds.
Therefore, the above estimates for the range of applicability of fluid dynamics may be extended to finite coupling strength, where
they lead to $|\bk| \lsim c T/(\eta/s)$ with the constant $c$ in the range $0.3 < c < 0.6$. In the strong coupling limit, in which $\eta/s = 1/4\pi$,
this reduces to the above estimates, but the range of validity of hydrodynamics will decrease with increasing dissipative
effects.  It will be very small in a weakly coupled system since $\eta/s$ takes parametrically large values $\eta/s \sim 1/\alpha_s^2$
 in perturbation theory. 

In summary, in a strongly coupled liquid we expect that higher order derivatives in the fluid dynamic description may be 
neglected for the description of wavevectors up to $|\bk| \sim 8\, T$, or so. Any increase in dissipation (i.e. any increase of $\eta/s$) reduces 
this range of validity of a fluid dynamic description. On the other hand, the inclusion of second-order terms in the fluid dynamic expansion
may extend this range of validity of fluid dynamics somewhat, while the expansion scheme is expected to break down always for sufficiently
large momenta. 

\section{Linear and nonlinear evolution}
\label{sec6}
In this section we solve numerically the time evolution for the spectrum and bispectrum derived in sect.\ \ref{correl},
starting from the initial conditions of the model of independent point sources introduced in section \ref{sec:InitialConditions}.

\begin{figure}[t]
\centering
\includegraphics[width=100mm,height=60mm]{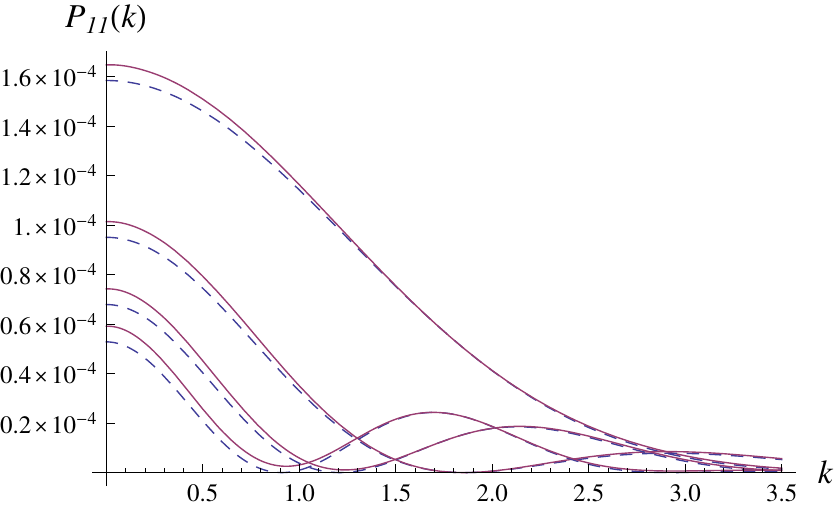}
\caption{
The density-density correlator $P_{11}(k)$ at $\mathcal{O}(\gamma^0)$ (blue, dashed line) and  $\mathcal{O}(\gamma^1)$ (purple, solid line).
The evolution was initialized at $\tau_0=1$ fm, for an initial kinematic viscosity $\nu_0=0.04$ fm corresponding to 
$\eta/s=1/(4\pi)$ and an initial temperature $T_0 \simeq 400$ MeV. 
The initial number density of sources
was $N/A=0.2$ fm$^{-2}$. Sources are chosen with a Gaussian profile of width $\sigma=0.4$ fm, that
amounts to cut-off initial fluctuations at momentum scales above 3 fm$^{-1}$. 
Results are shown for times 
$t=1.5,\, 2.5,\, 3.5,\, 4.5$ fm (curves from top right to bottom left). (All quantities measured in fm.)}
\label{p11evol}
\end{figure}

\subsection{Spectrum}
To solve for the spectrum at leading order, which corresponds to a linear time evolution, 
we can simply use the propagator obtained in sect.\ \ref{subsec:LinearPropagator}. 
For the initial conditions specified in eqn.\ \eqref{eq:InitialSpectrum}, we find
\begin{equation}
P_{ab}(k,t) {\big |}_\text{LO} = g_{a1}(k,t,t_0)\,  g_{b1}(k,t,t_0) \frac{1}{(2\pi)^2}\frac{A}{16 N} e^{-\sigma^2 k^2}.
\label{pablo} \end{equation}
At next-to-leading order, this expression gets supplemented by a term that involves 
the initial bispectrum of eq.\ \eqref{threepg} and a vertex term $\sim \gamma$. We obtain
\begin{equation}
\begin{split}
P_{ab}(k,t) {\big |}_\text{NLO} = & P_{ab}(k,t) {\big |}_\text{L.O.}  + \\
& \int_{t_0}^t dt^\prime \int d^2 q \left[ g_{ae}(k,t,t_0) g_{b1}(k,t,t_0) + g_{a1}(k,t,t_0) g_{be}(k,t,t_0) \right]  \\
&\times \gamma_{ecd}(-\bq,\bq-\bk) g_{c1}(q,t,t_0) g_{d1}(|\bq-\bk|,t,t_0) \frac{1}{(2\pi)^4} \frac{A^2}{64 N^2} e^{-\sigma^2(k^2+q^2-\bq \bk)}.
\end{split}
\label{pabnlo} \end{equation}
Higher orders are suppressed by additional powers of $1/N$.

%

\begin{figure}[t]
\begin{minipage}{72mm}
\includegraphics[width=\linewidth,height=50mm]{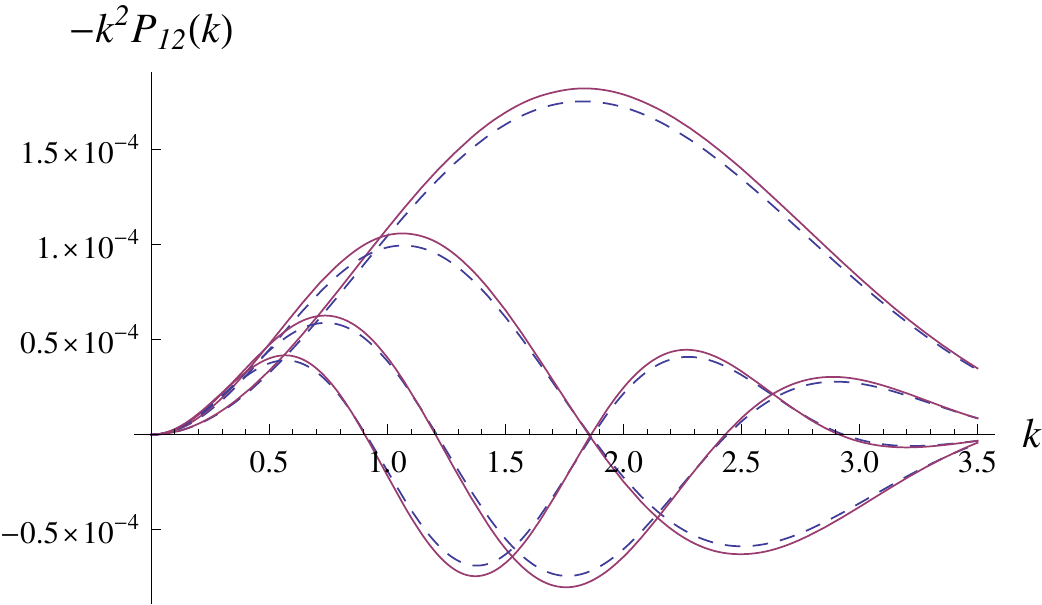}
\caption{
The correlation $-k^2 P_{12}(k)$ between density and velocity divergence at $\mathcal{O}(\gamma^0)$ (blue, dashed line) and  $\mathcal{O}(\gamma^1)$ (purple, solid line),
for  the same parameter choices as in fig. \ref{p11evol}. (All quantities measured in fm.) }
\label{p12evol}
\end{minipage}
\hfil
\begin{minipage}{72mm}
\includegraphics[width=\linewidth,height=50mm]{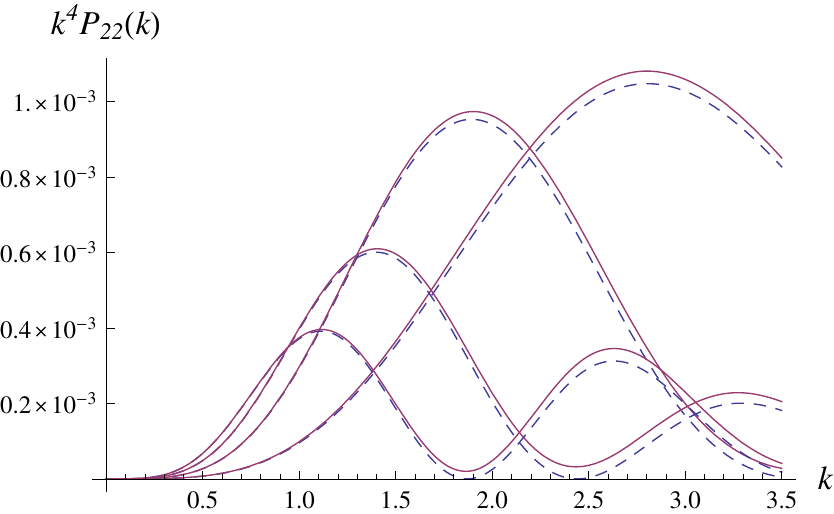}
\caption{
The correlation $k^4 P_{22}(k)$ between velocity divergence and velocity divergence 
 at $\mathcal{O}(\gamma^0)$ (blue, dashed line) and  $\mathcal{O}(\gamma^1)$ (purple, solid line),
for  the same parameter choices as in fig. \ref{p11evol}. (All quantities measured in fm.)}
\label{p22evol}
\end{minipage}
\end{figure}

The time dependence of the components $P_{11}(k,t)$ , 
$ -k^2 P_{12}(k,t)$  and  $ k^4 P_{22}(k,t)$  is shown in figs. \ref{p11evol}-\ref{p22evol}. The fluid dynamic evolution was initialized at $\tau_0=1$ fm, for an initial kinematic viscosity $\nu_0=0.04$ fm corresponding to 
$\eta/s=1/(4\pi)$ and an initial temperature $T_0 \simeq 400$ MeV  (see eq. (\ref{eq:Bjorkennutau})). The initial number density of sources
was $N/A=0.2$ fm$^{-2}$. Sources are chosen with a Gaussian profile of width $\sigma=0.4$ fm (see eq. (\ref{fourier})), that
amounts to cut-off initial fluctuations at momentum scales above 3 fm$^{-1}$ (see fig. \ref{hB04}). Since the velocity divergence $\theta$ and the field $f$ 
are related through $\tth_\bk=-k^2 \tf_\bk$, these spectra correspond to $d$-$d$, $d$-$\theta$ and $\theta$-$\theta$ correlations, respectively,
that is the self- and cross-correlations between the logarithmic energy density $d$ and the velocity divergence $\theta$. As seen from 
figs. \ref{p12evol}, \ref{p22evol}, velocity correlations have already built up at $t=1.5$ fm, even though the initial velocity
field is assumed to vanish at $t_0=0.75$ fm. 
The magnitude of all spectra diminishes at late times as a result of the longitudinal expansion of the
Bjorken background and the effect of viscosity. 

Figs. \ref{p11evol}-\ref{p22evol} show the development of oscillatory behavior which corresponds to the
appearance of acoustic waves induced by the strong pressure gradients in the relativistic fluid. The location of the minima is determined by the structure of the propagator that we discussed already in subsection \ref{subsec:LinearPropagator} in the context of equations (\ref{dbk}) and (\ref{fdk}). In particular, the argument of the Bessel functions in these propagators involves the combination $k\, t^{3/4}$, and this explains the continuous shift of the minima of the spectra towards smaller values for increasing time. As seen from figs. \ref{p11evol}-\ref{p22evol}, this behaviour is generic for all fields and it continues up to the time when the plasma is so dilute that the fluid description breaks down. Therefore, the spectra shown here are particularly suited for a discussion of how the spectrum evolves with time to a form where it is dominated by longer wavelengths. This is relevant for addressing one of the central topics in the fluid phenomenology of heavy-ion collisions, namely how  the spectrum of fluctuations at decoupling (that is experimentally accessible through final state hadrons) is related to the spectrum of initial fluctuations. There is a close analogy between this question and the question of how the power spectrum of acoustic oscillations at the time of recombination is fixed by the dynamics of microscopic interactions in cosmological scenarios. Access to the latter is provided by the spectrum of temperature fluctuations in the CMB and by the baryonic acoustic oscillations (BAOs) seen in the large scale structure of the Universe. 

Figs.  \ref{p11evol}-\ref{p22evol} also illustrate clearly that nonlinearities contribute only a small correction to the final form of the 
spectrum for the parameter values favored by the phenomenology of heavy-ion collisions. In the above figures, this conclusion can
be drawn from a comparison of leading $\mathcal{O}(\gamma^0)$ and next-to-leading $\mathcal{O}(\gamma^1)$ terms. The same
conclusion is also supported from the analysis of the $\mathcal{O}(\gamma^2)$ backreaction of the initial spectrum that we shall
discuss in the following. This is numerical evidence that a perturbative expasion of the spectrum and bispectrum in powers
of $\gamma$ has good convergence properties.

\begin{figure}[t]
\centering
\includegraphics[width=100mm,height=60mm]{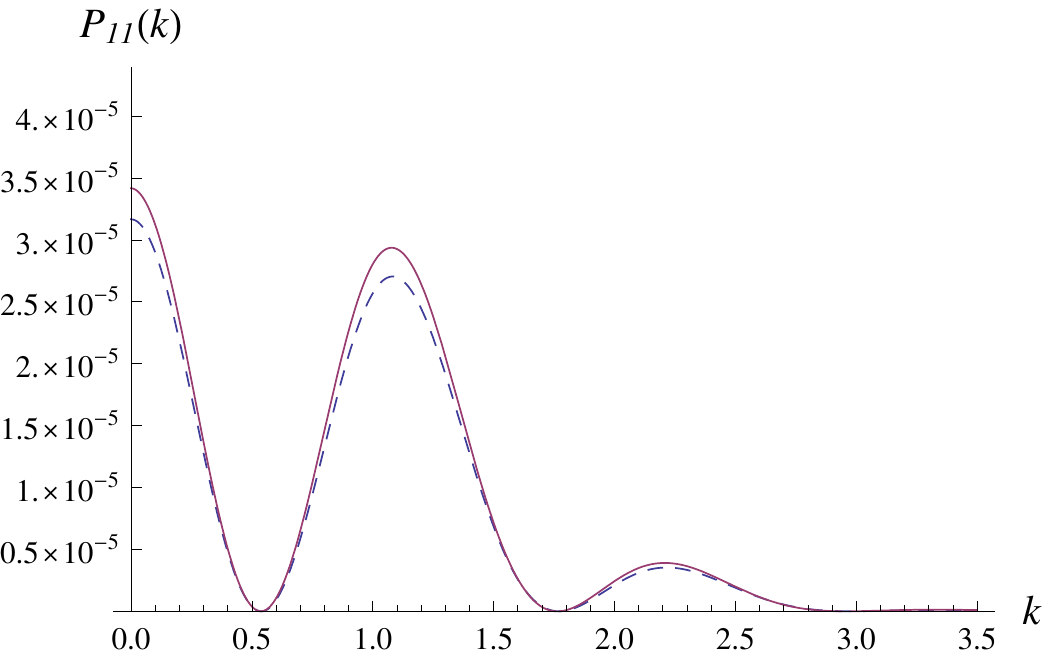}
\caption{
The spectrum $P_{11}(k)$ at $\mathcal{O}(\gamma^0)$ (blue, dashed line) and  $\mathcal{O}(\gamma^1)$ (purple, solid line) 
for the same parameters as in fig. \ref{p11evol}, but evolved up to time $t=7.5$ fm. (All quantities measured in fm.)}
\label{p11a}
\end{figure}

\begin{figure}[h]
\begin{minipage}{60mm}
\includegraphics[width=75mm,height=50mm]{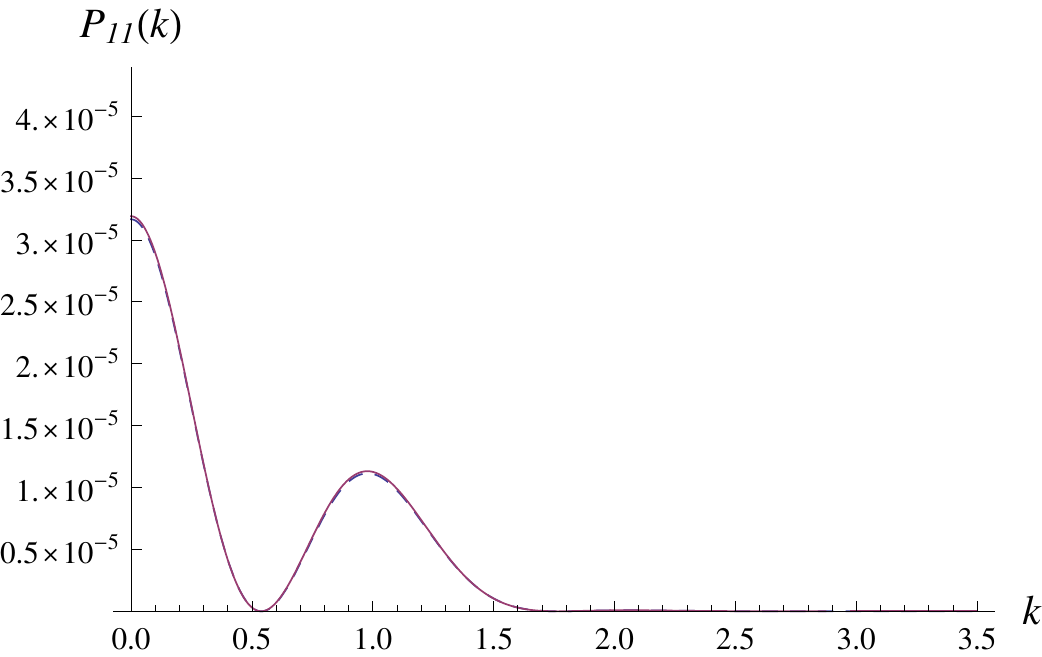}
\caption{
Same as fig.~\ref{p11a}, but for $\sigma=1$ fm. (All quantities measured in fm.)}
\label{p11b}
\end{minipage}
\hfil
\begin{minipage}{60mm}
\includegraphics[width=75mm,height=50mm]{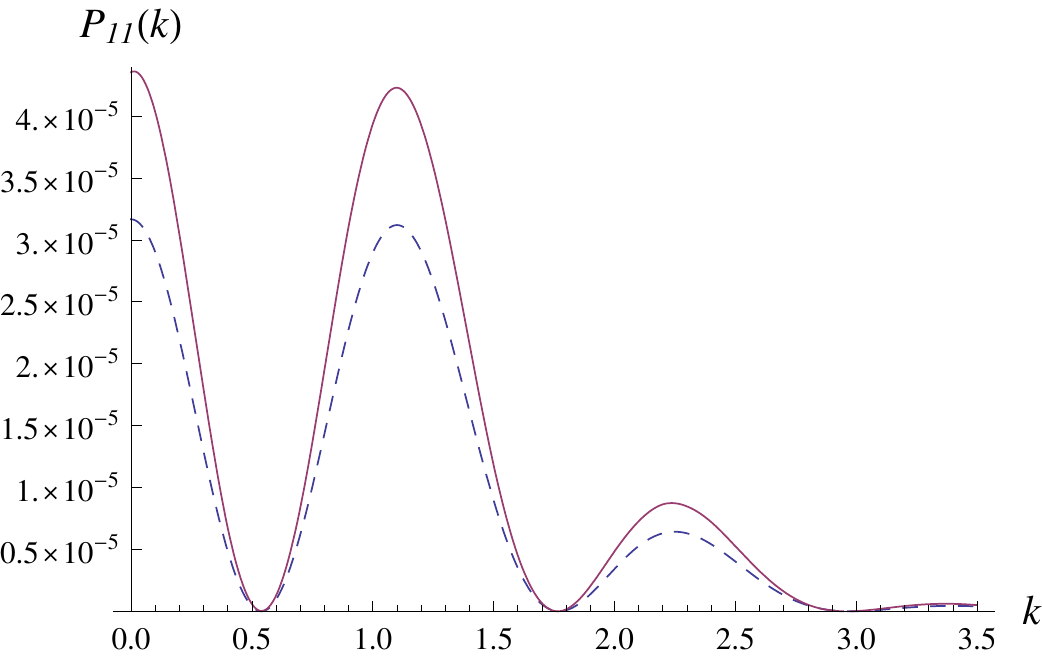}
\caption{
Same as fig.~\ref{p11a}, but for $\sigma=0.2$ fm. (All quantities measured in fm.)}
\label{p11c}
\end{minipage}
\end{figure}

We conclude this section by discussing in more detail the dependence of the numerical results on the parameters chosen. 
Figs. \ref{p11a} will serve as our reference, showing the density-density correlator for the same parameter choices as in
figs. \ref{p11evol}, but now evolved up to the later time $t=7.5$ fm. We compare fig. \ref{p11a} to figs.~\ref{p11b}-\ref{p11c} 
to illustrate how the spectrum depends on the maximal momentum scale of fluctuations in the initial conditions 
that can controlled be in our model  through the choice of the Gaussian width $\sigma$. 
For larger $\sigma$ the linear spectrum is suppressed at nonzero values of $k$, as the corresponding modes are eliminated from the 
initial spectrum. Moreover, the nonlinear corrections are also suppressed because the loop corrections receive contributions only from a
short range of low-momentum modes. As a consequence, nonlinear corrections to $P_{11}(k)$ become negligible, see  fig.~\ref{p11b}.
For small $\sigma$, a larger number of oscillations is visible in the spectrum, with large amplitudes
relative to the large-$\sigma$ case, and with somewhat larger nonlinear corrections (see fig.~\ref{p11c}). We observe that while all these dependencies may
have been expected on general grounds, their numerical size is sufficiently small to fully support the conclusions reached above. 

\begin{figure}[t]
\centering
\includegraphics[width=100mm,height=60mm]{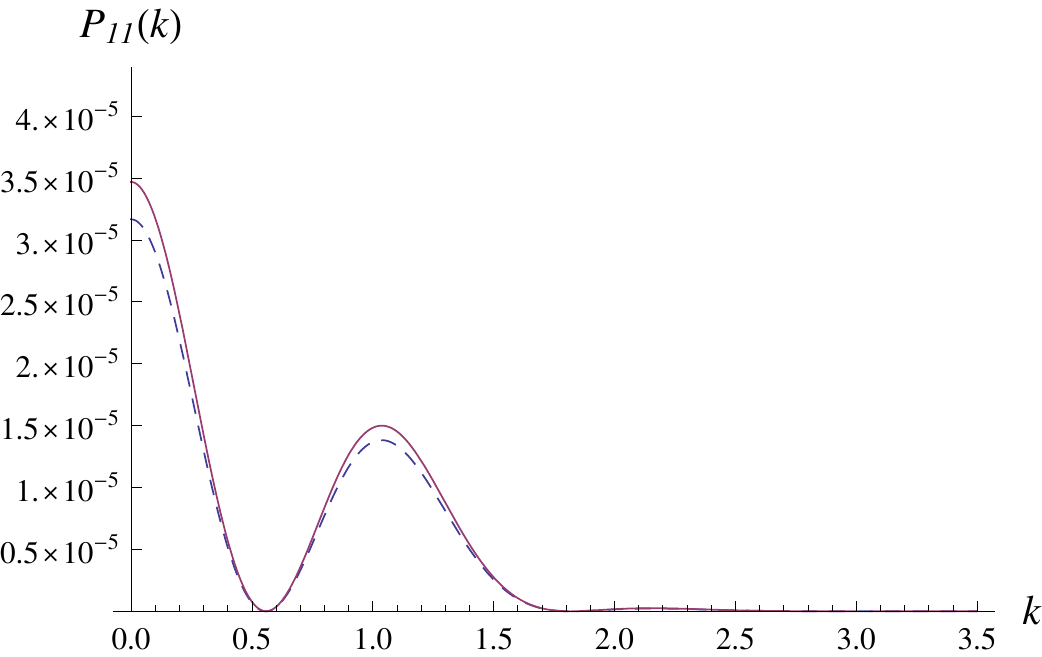}
\caption{
{$P_{11}(k)$ at $\mathcal{O}(\gamma^0)$ (blue, dashed line) and  $\mathcal{O}(\gamma^1)$ (purple, solid line) 
for  $\tau_0=1$ fm, $\nu_0=0.1$ fm, $\sigma=0.4$ fm at $t=7.5$ fm. (All quantities measured in fm.)}}
\label{p11f}
\end{figure}

We finally consider variations in kinematic viscosity by comparing the case of fig. \ref{p11a} corresponding to a minimal shear viscosity,
to the case of fig. \ref{p11f}, where shear viscosity is a factor 2.5 higher. Again, as expected,
viscous effects result in the suppression of the high-momentum modes, which are continuously depleted with time. 
As a result, at $t=7.5$ fm the second (third) peak of the acoustic oscillations in the spectrum are suppressed (strongly suppressed)
for $\nu=0.1$ fm relative to the $\nu=0.04$ fm case.  If such an analysis could be elevated to more realistic models of heavy-ion collisions,
the study of the relative sizes of the secondary peaks of the acoustic oscillations could provide a precision tool for further narrowing
uncertainties in the extraction of $\eta/s$. As an aside, we note the analogy between this observation and the current phenomenological analysis
of CMB fluctuations, in which the location and strength of peaks in the power spectrum are linked to material peroperties of the Universe.

\subsection{Bispectrum}
We next turn to the time evolution of the bispectrum. If the trispectrum is neglected, as in eq.\ \eqref{formalsol2}, 
there are two main contributions to the bispectrum. The leading-order one comes from the linear evolution of the initial bispectrum.  
For the initial condition of eq.\ \eqref{threepg}, it is of the form 
\begin{equation}
B_{abc}(k,q,p;t) {\big |}_\text{LO} = g_{a1}(k,t,t_0)\, g_{b1}(q,t,t_0) \, g_{c1}(p,t,t_0) \frac{1}{(2\pi)^3}\frac{A^2}{64 N^2} e^{-\frac{\sigma^2}{2} (k^2+q^2+p^2)}.
\end{equation}
At next-to-leading order, this gets supplemented by a term that is independent of the initial bispectrum and leads to a non-trivial bispectrum at late 
times,  even if $B_{abc} = 0 $ at $t=t_0$. The full expression is of the form
\begin{equation}
\begin{split} 
& B_{abc}(k,q,p;t) {\big |}_\text{NLO} = B_{abc}(k,q,p;t) {\big |}_\text{LO} \\
& + 2 \int_{t_0}^t dt^\prime {\Big [} g_{ad}(k,t,t^\prime) g_{b1}(q,t,t_0) g_{c1}(p,t,t_0) \gamma_{dgh}(\bq,\bp)g_{g1}(q,t^\prime,t_0) g_{h1}(p,t^\prime,t_0) \\
& \quad \times \frac{1}{(2\pi)^4} \frac{A^2}{256 N^2} e^{-\sigma^2(q^2+p^2)} + \text{two permutations}\; {\Big ]}. 
\end{split}
\end{equation}
The contribution from the nonlinear interactions does not factorize in general. 
Note that here the leading and next-to-leading terms are of the same order in $A/N$.

\begin{figure}[h]
\begin{minipage}{70mm}
\includegraphics[width=75mm,height=50mm]{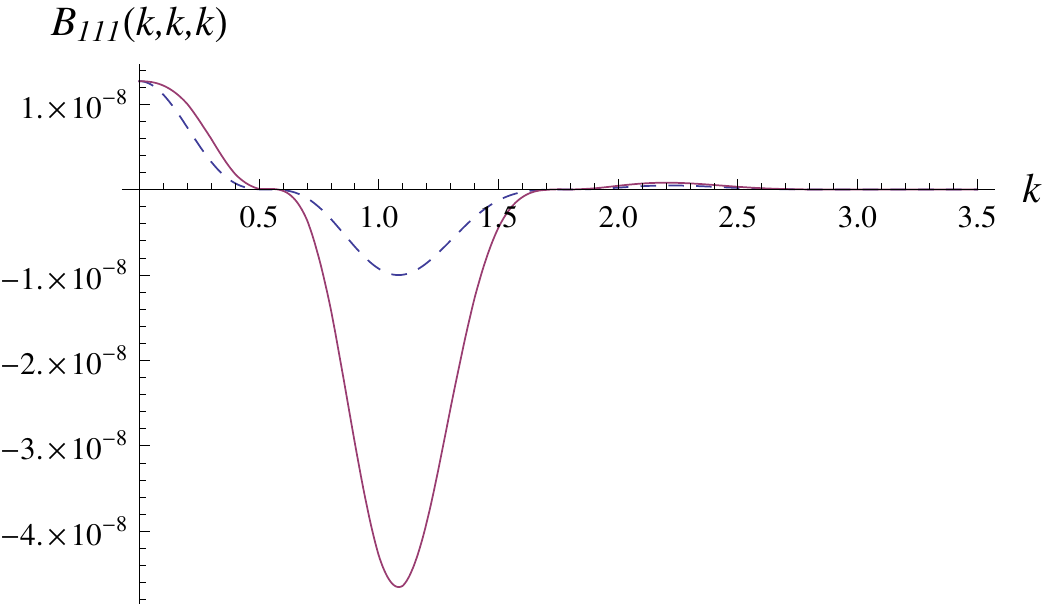}
\caption{
$B_{111}(k,k,k)$ at $\mathcal{O}(\gamma^0)$ (blue, dashed line) and  $\mathcal{O}(\gamma^1)$ (purple, solid line) 
for  $\tau_0=1$ fm, $\nu_0=0.04$ fm, $\sigma=0.4$ fm at $t=7.5$ fm. (All quantities measured in fm.)}
\label{b111a}
\end{minipage}
\hfil
\begin{minipage}{70mm}
\includegraphics[width=75mm,height=50mm]{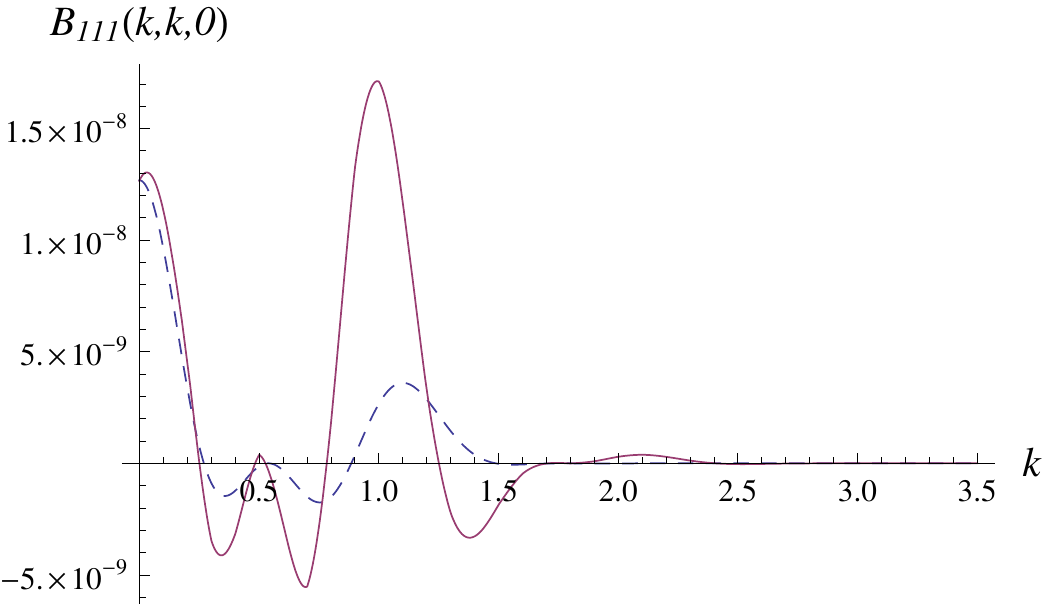} 
\caption{$B_{111}(k,k,0)$ at $\mathcal{O}(\gamma^0)$ (blue, dashed line) and  $\mathcal{O}(\gamma^1)$ (purple, solid line) 
for  $\tau_0=1$ fm, $\nu_0=0.04$ fm, $\sigma=0.4$ fm at $t=7.5$ fm. (All quantities measured in fm.)}
\label{b111b}
\end{minipage}
\end{figure}

In fig. \ref{b111a} we depict the $O(\gamma^0)$ and $O(\gamma^1)$ contributions to the 
111-component of the bispectrum at $t=7.5$ fm for momenta that form an equilateral triangle, so that $k=q=p$.
We observe that the linear evolution results in a spectrum with maxima and minima that reflect the oscillatory form of the propagator at
late times. Contrary to the spectrum discussed in the previous subsection, the $\mathcal{O}(\gamma^1)$ 
correction to the bispectrum is significant and overwhelms the linear contribution. The nonlinear effects tend to further enhance the 
bispectrum in the momentum ranges in which the linear evolution has generated significant deviations from zero.
While fig. \ref{b111a} shows the bispectrum for a particular choice $k=q=p$ of momenta, our finding of a large $O(\gamma^1)$
contribution persists also for other choices of arguments, such as the one shown in fig. \ref{b111b}. These 
results indicate that, for our model, the spectrum is the dominant quantity determining the form of the
bispectrum at late times. The initial condition for the bispectrum plays a secondary role. Since the form of the spectrum 
affects the bispectrum only at $O(\gamma^1)$, the large $O(\gamma^1)$ corrections seen in  fig. \ref{b111a}
and  fig. \ref{b111b} do not necessarily indicate the break-down of a perturbative approach but they rather
indicate that a qualitatively novel physics effect enters the bispectrum at $O(\gamma^1)$. However, the 
consistency of the perturbative approach requires then that higher order corrections remain small compared
to the $O(\gamma^0)$ and $O(\gamma^1)$ contributions.  It is important, therefore, to examine the
$\mathcal{O}(\gamma^2)$ correction to the spectrum, to which we turn now.


\subsection{$\mathcal{O}(\gamma^2)$ corrections and the backreaction}
\label{sec6.3}

In the following numerical exploration of $\mathcal{O}(\gamma^2)$ corrections to the spectrum, we limit ourselves again to the particular case
$P_{11}(\bk,t)$ at $\bk=0$. This simplifies the numerical task of solving eqs.~(\ref{formalsol1}), (\ref{formalsol2}), and it allows for some analytical 
insight. To $\mathcal{O}(\gamma^2)$, the solution of eq.~(\ref{formalsol1}) for $P_{ab}(\bk,t)$ involves a nonlinear backreaction of the
spectrum onto itself, and the analysis of $P_{11}(\bk=0,t)$ allows us to study this effect. 

\begin{figure}[t]
\begin{minipage}{60mm}
\includegraphics[width=75mm,height=50mm]{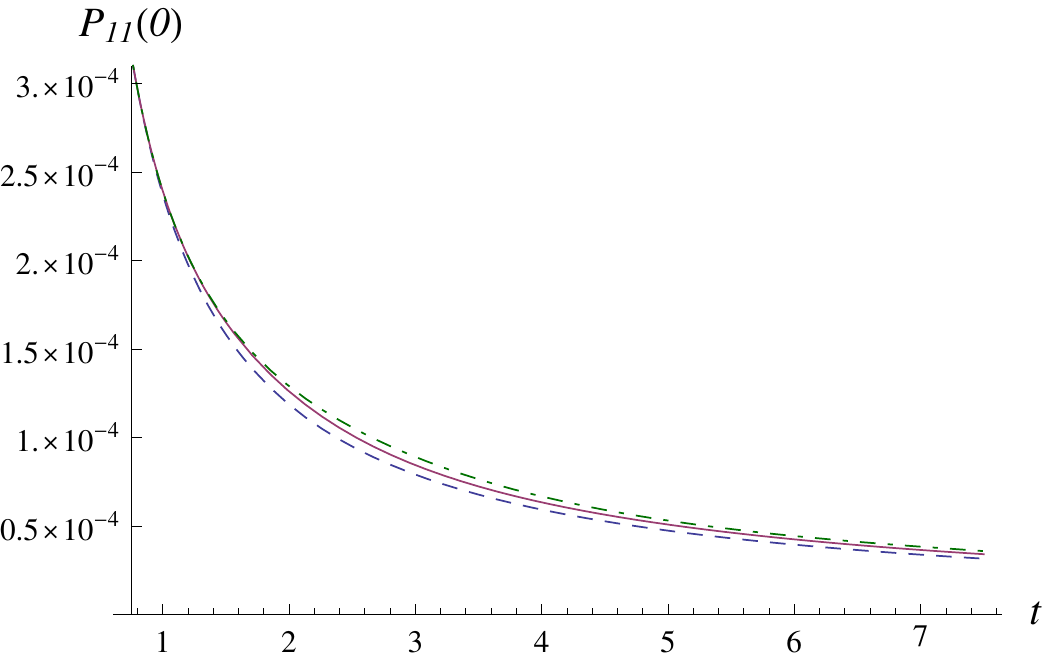}
\caption{
{ $P_{11}(0)$ at $\mathcal{O}(\gamma^0)$ (blue, dashed line),
$\mathcal{O}(\gamma^1)$ (purple, solid line) and 
$\mathcal{O}(\gamma^2)$ (green, dot-dashed line) as a function of $t$, for 
 $\tau_0=1$ fm, $\nu_0=0.04$ fm, $\sigma=0.4$ fm. (All quantities measured in fm.) }}
\label{p110}
\end{minipage}
\hfil
\begin{minipage}{60mm}
\includegraphics[width=75mm,height=50mm]{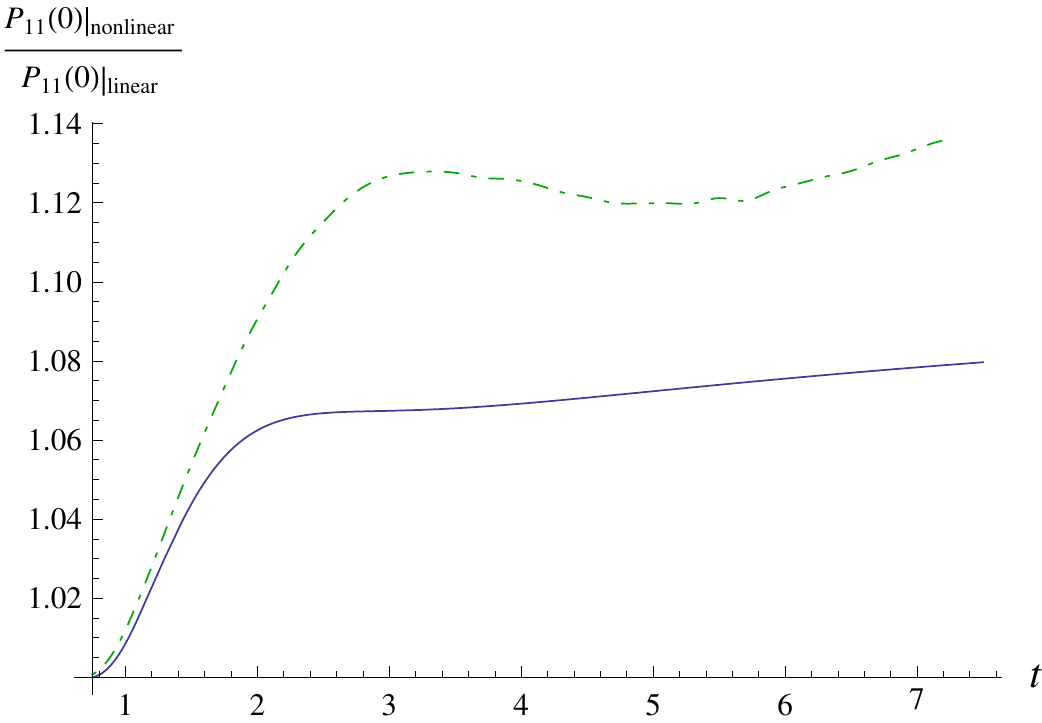}
\caption{
{ The ratio $P_{11}(0)|_{nonlinear}/ P_{11}(0)|_{linear}$ 
at order $\mathcal{O}(\gamma^1)$ (purple, solid line) and 
$\mathcal{O}(\gamma^2)$ (green, dot-dashed line) as a function of $t$, for 
 $\tau_0=1$ fm, $\nu_0=0.04$ fm, $\sigma=0.4$ fm. (All quantities measured in fm.) }}
\label{ratio}
\end{minipage}
\end{figure}

In fig. \ref{p110}, the time evolution of the 11-component of the spectrum for vanishing momentum is shown in three 
different approximations: linear,  $\mathcal{O}(\gamma^1)$ and  $\mathcal{O}(\gamma^2)$. 
All three lines are very close to each other, which indicates that the nonlinear corrections are small. This feature persists
also for other phenomenologically acceptable values of the parameters of the model, such as the ones we considered in 
the previous subsection. fig. \ref{ratio} shows the corresponding ratios of $P_{11}(\bk=0,t)$ evaluated at $\mathcal{O}(\gamma^1)$ 
and  $\mathcal{O}(\gamma^2)$ respectively and divided by the leading order contribution. This figure allows one to identify better
the time scale up to which nonlinearities develop, as well as the relative size of $\mathcal{O}(\gamma^1)$ 
and  $\mathcal{O}(\gamma^2)$ corrections. In particular, we see that for the parameter choices made for fig. \ref{ratio}, 
nonlinearities contribute to a growing deviation from the linear order up to a time $t \simeq 3-4$ fm, while the ratio to the linear
contribution remains almost unchanged for later times. This may be understood qualitatively by observing that the bispectrum 
contributes significantly to $P_{11}(\bk=0,t)$ in (\ref{funh}) only in the thin integration region $t'-t_0 \lsim 1/(\nu_0 q^2)$ and that
the most significant contributions comes from the highest momenta allowed by the cut-off in the initial conditions, $q \sim 1/\sigma$.
The resulting simple estimate $t_{\rm turn} \simeq \sigma^2/ \nu_0$ provides indeed a good order of magnitude estimate for the
time $t_{\rm turn}$ at which the curves in fig. \ref{ratio} stop rising. Numerically, the nonlinear effects amount to roughly 10\% of 
the final value of the spectrum, and they develop up to the time $O(2)$ fm/c after the onset of the hydrodynamic regime.
The $\mathcal{O}(\gamma^2)$ correction is of comparable size but not larger than the $\mathcal{O}(\gamma^1)$ correction. 

\begin{figure}[t]
\centering
\includegraphics[width=100mm,height=60mm]{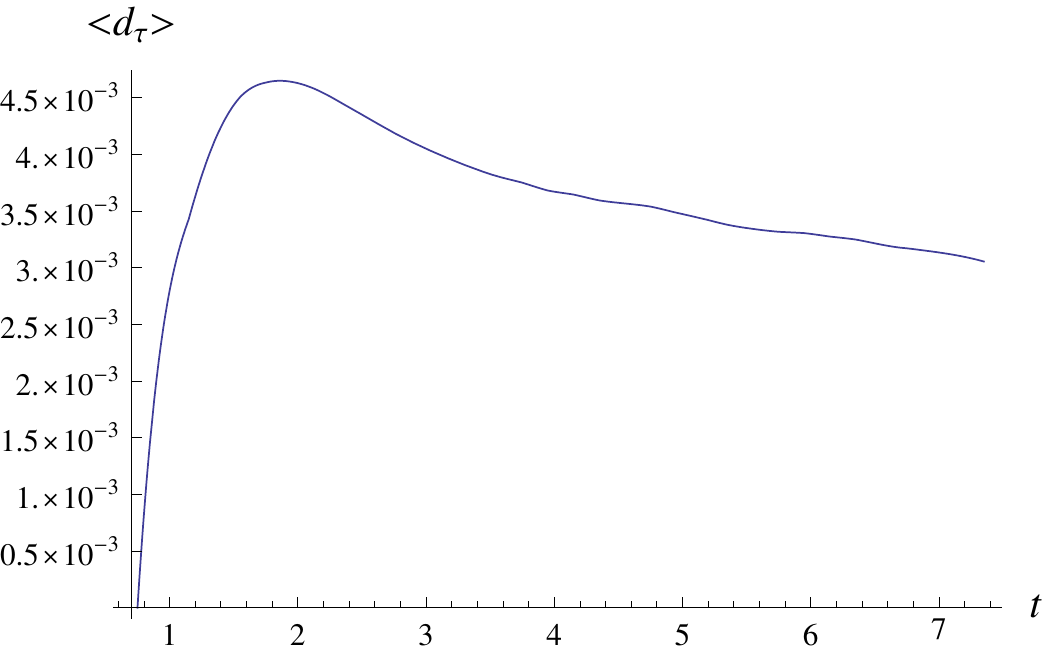}
\caption{
{ Backreaction of the spectrum on the expectation value of the density field, for  $\tau_0=1$ fm, $\nu_0=0.04$ fm, $\sigma=0.4$ fm }. (All quantities measured in fm.)}
\label{tadpole}
\end{figure}

We discuss finally the development of a nonzero expectation value for the density. This is a 
qualitatively novel feature that arises due to the backreaction of the spectrum. 
Translational invariance requires that nonzero Fourier modes of the fields do not develop expectation values. However, the 
$\bk=0$ mode can become nonzero during the time evolution of the system
because of nonlinear effects, even if one sets intially $\bar{\phi}_a(t_0) = 0$  through an appropriate field definition at the initial time,
as we did in section~\ref{sec:defExpValuesSpectra}. 
As we have argued already in section \ref{sec:defExpValuesSpectra}, the expectation value of $f(t,\bx)$ has no physical meaning and can always 
be set to zero. On the other side, the expectation value $\bar\phi_1(t) = \bar d_\tau(t)$ is physical. 
In particular, 
eq.\ (\ref{formalsol1e}) gives
\be
\bar d_\tau (t)= \int_{t_0}^t dt^\prime  \int d^2 q \, g_{ab}({\bf 0}\,,t,t^\prime) 
\left( -q^2  P_{12} (\bq,t') +  \frac{4}{9} \nu_0 \; q^4  P_{22} (\bq,t') 
\right).
\label{deltatau} 
\ee
The last term, which is positive semi-definite, describes the effect of dissipation, while the previous one corresponds to a nonlinear convection effect. 
Physically, eq.~(\ref{deltatau}) characterizes how the dissipation and convection of modes at finite momentum affects the spatially averaged background field.
In fig. \ref{tadpole} we depict this time evolution of the $\bk=0$ mode of the density field. In configuration space, this mode corresponds to
the density expectation value. At the initial time, our choice of initial conditions enforces a vanishing value, $\bar d_\tau (t_0) =0$. 
However, the subsequent evolution leads to a quick deviation from zero, developing within a short time scale of $\sim 1$ fm, similarly to 
what we observed in fig. \ref{ratio}. At late times, the density drops  slowly because of the expansion of the Bjorken background. 
As we discussed in subsection \ref{sec:defExpValuesSpectra}, an appropriate time-dependent field redefinition, that would absorb this effect,  is required  for
a precise statistical analysis of the late-time fluctuations around the shifted background.

\section{Discussion}

In this work, we have calculated the time evolution of two- and three-point functions (a.k.a. spectrum and bispectrum) of fluid
dynamic fields for a simplified fluid model of ultra-relativistic heavy-ion collisions, the so-called Bjorken model. The analysis
of spectra and bispectra is a standard tool in the study of cosmological perturbations, while the fluid dynamic formulation of 
perturbations is a recent development in the phenomenology of ultra-relativistic heavy-ion collisions. Even though the structural analogies between
fluid dynamic applications in cosmology and in heavy-ion physics have been remarked on repeatedly, the present work is, to the best of our 
knowledge, the first in which results for spectra and  bispectra are presented for a model of heavy-ion collisions. Here, we summarize our main findings:

After having introduced a background-fluctuation splitting of fluid dynamic fields for the Bjorken model in section~\ref{sec2},
we found in section~\ref{correl} that the evolution equations for the spectrum and bispectrum of fluctuations can be written in a
perturbative expansion that is of the same form as the expansion used for the study of cosmological perturbations.  In section~\ref{sec:InitialConditions},
we provided a simple but sufficiently generic model for the {\it initial} spectrum and bispectrum of fluctuations, through which this perturbative
dynamics can be initialized. 
Within this setting, which is technically different from other fluid dynamic formulations of heavy-ion collisions, we
analyzed the range of wavelengths up to which a fluid dynamic evolution is applicable, as well as the range of validity of a perturbative expansion
for physically relevant model parameters. To this end, we discussed in detail the UV sensitivity of the fluid dynamic evolution. We 
observed that, depending on the choice of independent fluid dynamic fields, the UV sensitivity seems to arise either from linear or from nonlinear
terms in the fluid dynamics (see subsection~\ref{sec:DerCouplingsUVProp}). We also found strong numerical evidence in support of the assumption that  a perturbative
expansion applies to the calculation of the spectra and bispectra of fluid dynamic fluctuations of realistic size (see section~\ref{sec6}). 

The correlated two- and three-point functions of fluid dynamic fields calculated here are not directly measurable in heavy-ion collisions. 
The reason is that we did not supplement the fluid dynamic evolution with a hadronization prescription, while the measurable momentum
correlations of hadrons provide only convoluted information about spatial correlators. However, the spectra and bispectra calculated here provide 
arguably the most direct information about how an initial spectrum of fluctuations evolves as a function of time and how perturbations dissipate (i.e. weaken in strength). Any attempt of constraining the scale and nature of fluctuations at the initial time $\tau_0$ via experimental measurements of fluctuations 
at freeze-out relies on understanding the dynamic evolution of the scale and size of fluctuations. 
We expect that the analysis of fluid dynamic spectra and bispectra will be useful to this end.

In a setup with realistic radial dependence of the background it is possible to obtain correlation functions of hadrons in momentum space from the correlation functions of fluid dynamic variables as discussed here. Ref.\ \cite{Floerchinger:2013hza} provides the formalism for this in terms of appropriate integrals over the freeze-out surface of the background solution, to linear order in the perturbations.

We finally point out that the dynamics and in particular the dissipation of fluctuations results in a backreaction that can be characterized by a growing,
spatially independent zero-mode on top of the unperturbed background field. We have explained in subsections~\ref{sec3.3} and~\ref{sec6.3} the physics 
encoded in this term and how it arises technically. While the present paper was motivated by the idea of adopting an analysis technique used in cosmology 
to a problem in heavy-ion physics, a companion paper discusses the physics of this backreaction effect in cosmology \cite{FTW2014}.

\medskip
\section*{Acknowledgments}
\medskip
We would like to thank M. Garny, D. Blas and M. Pietroni for useful discussions.  
The work of N.T. has been supported in part 
by the European Commission under the ERC Advanced Grant BSMOXFORD 228169.
The work of N.B. and N.T. has been co-financed by the European Union (European Social Fund ESF) and Greek national 
funds through the Operational Program ``Education and Lifelong Learning" of the National Strategic Reference 
Framework (NSRF) - Research Funding Program: ``THALIS. Investing in the society of knowledge through the 
European Social Fund''. 


\end{document}